\documentclass[final,5p,times,twocolumn]{elsarticle}

\usepackage[utf8]{inputenc}
\usepackage{geometry}
\usepackage{amsmath,amssymb,amsfonts} % standard math symbols/fonts
\usepackage{algorithmic} % ?
\usepackage{graphicx} % ?
\usepackage{textcomp} % ?
\usepackage{xcolor} % ?
\usepackage{hyperref} % make clickable links to citation/figure/table references 
\usepackage{hypernat}
\usepackage{natbib} % ?
\usepackage[parfill]{parskip} % no indent, space between paragraphs
\usepackage{caption,annotate-equations}

\usepackage{lineno}

\DeclareMathOperator{\sgn}{sgn}

\journal{}

\begin{document}

\begin{frontmatter}

%% Title, authors and addresses

%% use the tnoteref command within \title for footnotes;
%% use the tnotetext command for theassociated footnote;
%% use the fnref command within \author or \address for footnotes;
%% use the fntext command for theassociated footnote;
%% use the corref command within \author for corresponding author footnotes;
%% use the cortext command for theassociated footnote;
%% use the ead command for the email address,
%% and the form \ead[url] for the home page:
%% \title{Title\tnoteref{label1}}
%% \tnotetext[label1]{}
%% \author{Name\corref{cor1}\fnref{label2}}
%% \ead{email address}
%% \ead[url]{home page}
%% \fntext[label2]{}
%% \cortext[cor1]{}
%% \affiliation{organization={},
%%             addressline={},
%%             city={},
%%             postcode={},
%%             state={},
%%             country={}}
%% \fntext[label3]{}

\title{A Tri-Level Optimization Model for Interdependent Infrastructure Network Resilience Against Compound Hazard Events}

%% use optional labels to link authors explicitly to addresses:
%% \author[label1,label2]{}
%% \affiliation[label1]{organization={},
%%             addressline={},
%%             city={},
%%             postcode={},
%%             state={},
%%             country={}}
%%
%% \affiliation[label2]{organization={},
%%             addressline={},
%%             city={},
%%             postcode={},
%%             state={},
%%             country={}}

\author[1]{Matthew R. Oster}
\author[1,3]{Ilya Amburg\corref{aaa}}
\author[1,2]{Samrat Chatterjee\corref{aaa}}
\author[4]{Daniel A. Eisenberg}
% \author[2]{Jack Watson}
% \author[1]{Daniel Corbiani}
% \author[1]{Jennifer Webster}
\author[1]{Dennis G. Thomas}
\author[1]{Feng Pan}
\author[2,1]{Auroop R. Ganguly}
% \author[1]{William B. Gattis}
% \author[1]{Kyle Haynie}

\affiliation[1]{organization={Pacific Northwest National Laboratory}, 
                % addressline={?},
                city={Richland},
                state={WA},
                postcode={99352},
                country={USA}}
                
\affiliation[2]{organization={Northeastern University}, 
                % addressline={?},
                city={Boston}, 
                state={MA},
                postcode={02115},
                country={USA}}
\affiliation[3]{organization={Rochester Institute of Technology}, 
                % addressline={?},
                city={Rochester}, 
                state={NY},
                postcode={14623},
                country={USA}}   
\affiliation[4]{organization={Naval Postgraduate School}, 
                % addressline={?},
                city={Monterey}, 
                state={CA},
                postcode={93943},
                country={USA}}  
% \date{} % blank, vs \date{\today}
\cortext[aaa]{Corresponding Authors: \{firstname.lastname\} @pnnl.gov}

\begin{abstract}
Resilient operation of interdependent infrastructures against compound hazard events is essential for maintaining societal well-being. To address consequence assessment challenges in this problem space, we propose a novel tri-level optimization model applied to a proof-of-concept case study with fuel distribution and transportation networks -- encompassing one realistic network; one fictitious, yet realistic network; as well as networks drawn from three synthetic distributions. Mathematically, our approach takes the form of a defender-attacker-defender (DAD) model—a multi-agent tri-level optimization, comprised of a defender, attacker, and an operator acting in sequence. Here, our notional operator may choose proxy actions to operate an interdependent system comprised of fuel terminals and gas stations (functioning as supplies) and a transportation network with traffic flow (functioning as demand) to minimize unmet demand at gas stations. A notional attacker aims to hypothetically disrupt normal operations by reducing supply at the supply terminals, and the notional defender aims to identify best proxy defense policy options which include hardening supply terminals or allowing alternative distribution methods such as trucking reserve supplies. We solve our DAD formulation at a metropolitan scale
and present practical defense policy insights against hypothetical compound hazards. We demonstrate the generalizability of our framework by presenting results for a realistic network; a fictitious, yet realistic network; as well as for three networks drawn from synthetic distributions. Additionally, we demonstrate the scalability of the framework by investigating runtime performance as a function of the network size. Steps for future research are also discussed.

\end{abstract}

% %%Graphical abstract
% \begin{graphicalabstract}
% \includegraphics[width=0.77\linewidth]{images/phases.png}\includegraphics[width=1.22\linewidth]{images/defense}
% \end{graphicalabstract}

% %%Research highlights
% \begin{highlights}
% \item gain practical insights against compound threats in interdependent networks
% \item defender-attacker-defender model solved using column-and-constraint generation
% \item gain insights through case studies and demonstrate scalability on realistic networks
% \end{highlights}

\begin{keyword}
%% keywords here, in the form: keyword \sep keyword
defender-attacker-defender \sep critical infrastructure \sep resilience \sep optimization \sep fuel \sep transportation

%% PACS codes here, in the form: \PACS code \sep code
% \PACS 0000 \sep 1111
%% MSC codes here, in the form: \MSC code \sep code
%% or \MSC[2008] code \sep code (2000 is the default)
% \MSC 0000 \sep 1111
\end{keyword}

\end{frontmatter}

%% \linenumbers

%%% INTRO %%%
\section{Introduction}
As societal well-being becomes increasingly reliant on connected and reliable operation of lifeline infrastructures, such as energy and transportation, there is a growing need for maintaining resilient system functionality against a spectrum of natural and man-made hazards~\cite{rinaldi2001identifying,wells2022modeling}. Disruptions of interdependent critical infrastructure systems may threaten the health, security, and economies on scales ranging from local to global~\cite{ouyang2014review}. As a result, assessing consequences of \textit{compound} natural and man-made hazard events (i.e., simultaneous or sequential with localized or widespread effects) on infrastructure operations is critical for development of mitigation options and practical response planning insights. This becomes increasingly challenging with operationally interdependent infrastructures as well as multiple stakeholder objectives and decision priorities.     

Resilience studies of critical infrastructure networks have utilized attacker-defender (AD) models for the past several decades~\cite{brown2006defending,an2011guards}. In such settings, the attacker and defender act in sequence and share an objective function, where the defender aims to minimize the disruption that the attacker seeks to maximize. However, while such bi-level approaches allow assessing the vulnerability of systems, they do not allow planning in advance to prepare against potential attacks.

To mitigate these shortcomings, defender-attacker-defender (DAD) models were proposed, which allow the defender to prepare before the attacker acts~\cite{alderson2011solving}. Such models not only enable studying the resilience of critical infrastructure networks, but also include prescribing actions that could be taken to improve resilience~\cite{alderson2014assessing,alderson2015operational}. See Eqn.~\ref{eqn:dad} for a typical form of the resulting objective. In this work, we utilize the DAD model because of its additional prescriptive capability. The model assumes that all actors (i.e., decision agents) have complete knowledge, share an objective (rendering their interactions a zero-sum game), act in sequence, and play only one round~\cite{oster2020}. Although the resulting problems are often NP-hard to solve even in stylized cases~\cite{smith2020survey}, many recent developments in computational approaches have made industrial-scale optimization possible. For example, the ``dualize-and-combine'' approach replaces the inner bi-level optimization problem with a single level optimization problem. Further, when the attacker is only allowed binary decisions, the problem can be cast as a mixed integer program (MIP), for which many off-the-shelf solvers exist. Although traditionally Benders decomposition-type approaches have been applied to these kinds of problems, the recently-developed column-and-constraint generation algorithm outperforms Benders decomposition, as it is guaranteed to converge in fewer iterations~\cite{zeng2013}. In addition, DAD models are amenable to incorporating uncertainty, though we do not focus on that aspect in this study~\cite{infanger1992monte}.

Critical infrastructures are increasingly interdependent~\cite{milanovic2017modeling,ouyang2014review,rinaldi2001identifying}, where one system depends on the functioning of others. For example, water distribution systems may depend on the power grid to operate water pumps and quality control instruments, transportation relies on the power grid for traffic regulation, and transportation relies on the fuel distribution network for supplying fuel demand. As a result, it is important for models of resilience in critical infrastructure to incorporate such interdependence~\cite{liu2018vulnerability,setola2009critical,stergiopoulos2016time,tsavdaroglou2018proposed,luiijf2021analysis,lam2018modeling,lauge2015critical,oliva2010agent,thompson2019interdependent,seppanen2018critical,rehak2016quantitative}. Further, real-life threats are often compound, with multiple attacks/disruptions happening simultaneously or in sequence~\cite{jackson2013resilience,cutter2018compound,zhang2021hypothesis}. While some work has been done in modelling resilience of interdependent systems under compound threats~\cite{kuc2020,yadav2020resilience, chatterjee2021applied}, the literature is relatively sparse on models capable of generating practical insight to prepare against compound threats in interdependent networks~\cite{wells2022modeling}. This work contributes to filling this gap by applying the DAD model in the presence of compound hazards, yielding practical defense preparation policies. 

Although the model and framework that we present is general, we focus on the scenario of interdependent fuel and transportation networks. While the literature has several works on this scenario, none apply the DAD model with multiple decision variables as presented here.  In particular, ~\cite{ibanez2010interdependencies} presents a single-level network optimization approach for analyzing failure in fuel-transport networks. ~\cite{beheshtian2017planning} models failure in the presence of hurricanes/flooding. ~\cite{kuc2020} presents a combo model for analyzing the fuel-transport networks. Their model, in turn, builds on work in~\cite{good2019operational,routley2020operational}. Furthermore, there is a wider range of literature analyzing only transportation networks~\cite{wang2022vulnerability,wang2023vulnerability}.

Prior research on resilience-centric infrastructure operations ranges from assessing system vulnerability and embedding resilience through multi-agent/multi-level optimization model variants~\cite{ottenburger2020novel,tiong2023two} (see, {\em e.g.}, Oster {\em et al.}~\cite{oster2020} for a brief overview, and Smith and Song~\cite{smith2020survey} for a more in-depth review), to understanding interdependencies between systems~\cite{kuc2020, ouyang2014review, alderson2015operational, liu2018vulnerability,setola2009critical,stergiopoulos2016time,tsavdaroglou2018proposed,luiijf2021analysis,lam2018modeling,lauge2015critical,oliva2010agent,thompson2019interdependent,seppanen2018critical}, to simulating cascading impacts through $n-k$ contingency analysis~\cite{sundar2018probabilistic}, to deep learning and probabilistic approaches~\cite{wang2021resilience,hossain2019framework}. However, as model granularity is refined or model size is increased, computing at scale can become limited. Fortunately, general heuristic techniques typically exist for approximating solutions and offline machine learning methods appear promising for speeding up existing optimization algorithms~\cite{misra2018}.

In this paper, we address consequence assessment challenges associated with resilient operation of interdependent infrastructures against compound hazard events by proposing a novel tri-level optimization model. We apply this model to a proof-of-concept case study with fuel distribution and transportation networks. We embed our problem in a DAD framework. Here, our notional operator operates an interdependent system comprised of fuel terminals and gas stations and a transportation network with traffic flow so as to minimize a combination of unmet demand at gas stations and travel time for customers on the network. The notional attacker aims to disrupt normal operations by reducing supply at the supply terminals and gas stations which may induce traffic congestion, and the notional defender aims to identify best proxy defense policy options which include hardening supply terminals or opening reserve nodes. 

Our contributions are as follows:
\begin{itemize}
    \item We generalize the operator's fuel and transportation optimization combo-model of Kuc \cite{kuc2020}, where we allow multiple transportation modes and multiple supply phases
    \item We embed our model within a DAD framework and show how to solve it with the column-and-constraint generation (CCG) algorithm~\cite{zeng2013}
    \item We focus on a subclass of our generic model, where defense policies include fuel terminal supply node hardening and reserve node opening, while supply attack scenarios are compound events in that terminal supplies and gas station supplies suffer weather and/or intentional disruptions, and finally the operator resolves to route tanker trucks to supply gas stations while customers are routed, balancing aggregate travel time and unmet system-wide demand
    \item We apply our model to fictitious, yet realistic fuel-transportation networks on the St. Thomas, U.S. Virgin Islands (USVI) and the realistic Anaheim, California transportation network -- generating practical decision-support insights
    \item We demonstrate the generalizability of our model to multiple use cases by generating results for the fictitious, yet realistic St. Thomas, USVI and realistic Anaheim networks, as well as for networks drawn from three synthetic distributions that are often used to model transportation networks
    \item We demonstrate the potential of our model to scale to larger scenarios by solving instances on progressively larger synthetic networks
    
\end{itemize}

%%% METHODOLOGY %%%
\section{Methodology}
To formulate our optimization problem as a DAD model, we require a few assumptions. In particular, we require that the three participating decision agents (i.e., \textit{defender}, \textit{attacker}, and \textit{operator}) make decisions in sequence, have complete information, share an objective function, and are certain of their individual effects downstream. In particular, the optimization problem takes on the following form:
\vspace{1cm}
\begin{equation}
\!\! (\textrm{DAD}) \quad \eqnmarkbox[blue]{p1}{\min_{w \in W}} \left\{ \eqnmarkbox[red]{p2}{\max_{x \in X}} \left\{ \eqnmarkbox[teal]{p3}{\min_{y \in Y(w, x)}} f(w, x, y) \right\} \right\} \label{eqn:dad}
\end{equation}\annotate[yshift=2em]{above}{p1}{Defender prepares}\annotate[yshift=-0.15em]{below}{p2}{Attacker acts}\annotate[yshift=1em]{above}{p3}{Operator manages}

\vspace{0.5cm}
where $f$ is a real-valued scalar map, $W = \{w \in \mathbb{R}^{\ell} \times \mathbb{Z}^{\ell'}: F(w) \geq 0 \}$ captures the feasible decision space of the defender, $F$ is a vector map, and $\ell, \ell' \geq 0$.  Further, $X = \{x \in \mathbb{R}^{m} \times \mathbb{Z}^{m'}: G(x) \geq 0 \}$ captures the feasible decision space of the attacker for any feasible defense $w \in W$, $G$ is a vector map, and $m, m' \geq 0$.  Similarly, and finally, $Y(w, x) = \{y \in \mathbb{R}^{n} \times \mathbb{Z}^{n'}: H(w, x, y) \geq 0\}$ captures the feasible response decision space of the defender for any $w \in W$ and $x \in X$, $H$ is a vector map, and $n, n' \geq 0$.

Once in this form, we solve our model with the column-and-constraint generation (CCG) algorithm of~\cite{zeng2013}, which is advantageous due to its dominance over the classical Benders decomposition. The CCG algorithm solves this tri-level problem by iterating between solving a master problem (MP) and a subproblem (SP), each providing tighter bounds on opposite ends of the true optimum that is eventually found.

In particular, the master problem will have the following form:
\begin{align}
(\textrm{MP}(I)) \quad \min_{\eta \in \mathbb{R}, w \in W, y^i \in Y(w, x^i), i \in I} & \eta \\
\textrm{subject to} \quad \eta \geq f(w, x^i, y^i) &\quad \forall i \in I,
\end{align}
where $I$ contains a subset of indices of all extremal (attack) points $x^i \in X$.  Solving $MP(I)$ for any such $I$ results in a best feasible defense $w \in W$ robust against the subset $I$, and thus yielding a lower bound to \eqref{eqn:dad}.

For the subproblem, we solve the following problem for a given defense vector $w \in W$:
\begin{align}
(\textrm{SP}(w)) \quad & \max_{x \in X} \min_{y \in Y(w, x)} f(w, x, y) \\
= & \max_{x \in X, z \in Z(w, x)} g(w, x, z),
\end{align}
which results in a best feasible attack $x \in X$ robust against the feasible defense $w$, and thus yielding an upper bound to \eqref{eqn:dad}.  Here the equality reducing the initial bilevel problem to a single-level one is possible if, as in our case, the inner-level formulation $\min_{y \in Y(w, x)} f(w, x, y)$ is a linear program for any given $w \in W, x \in X$, and strong duality is applied (yielding an equivalent dual problem $\max_{z \in Z(w, x)} g(w, x, z)$, though this may initially include bilinear terms in $w, x, z$).  Next, we specify and reformulate our master and subproblems as mixed-integer programs.

\section{Model Formulation}
We embed the DAD model in a simulation setting where a hypothetical event may disrupt system components (e.g., supply locations), encoding the effects of either natural or man-made compound events. We consider the defender to harden these potential avenues of disruption, or reduce disruption effects on the system. Tables~\ref{tab:units}, \ref{tab:sets}, \ref{tab:constants}, \ref{tab:variables}, and \ref{tab:params} present the units, sets, constants, variables, and parameters, respectively, in our DAD model described below. 

\begin{table}[htbp!]
  \begin{center}
    \caption{Units}
    \label{tab:units}
    \begin{tabular}{l|l}
      \textbf{Unit} & \textbf{Description}\\
      \hline
      \# & unitless quantity \\
      bbl & barrels of fuel \\
      u & (non-standard) vehicles \\
      v & (standard) vehicles \\
      mi & miles \\
      \$ & US dollars \\
      x/y & x per y (for any units x, y) \\
      x-y & x spanning y (e.g., power $\xrightarrow{}$ energy Mw-h) \\
      \hline
    \end{tabular}
  \end{center}
\end{table}

\begin{table}[htbp!]
  \begin{center}
    \caption{Sets}
    \label{tab:sets}
    \begin{tabular}{l|l}
      \textbf{Set} & \textbf{Description}\\
      \hline
      $ M $ & set of transportation networks or modes \\
      $ P $ & set of supply phase indices \\
      & \quad $(P =\{1, 2, \ldots, n_P\})$ \\
      $L$ & set of indices for approximation of BPR function \\
      & \quad $(L = \{1, 2, \ldots, n_L\})$ \\
      $ C_{mp} $ & set of carrier types in mode $m \!\in\! M\!$, during phase $p \!\in\! P$ \\
      $ V_{mp} $ & set of carrier-common nodes for $m \!\in\! M, p \in P$ \\
      $ A_{mp} $ & set of carrier-common directed arcs for $m \!\in\! M, p \in P$ \\
      $S_{mp}$ & subset of attacked or defended nodes in $V_{mp}$, $m \!\!\in\!\! M\!, \!p \!\!\in\!\! P$ \\
      $R_{mp}$ & subset of reserve supply nodes in $V_{mp}$, $m \!\!\in\!\! M\!, \!p \!\!\in\!\! P$ \\
      $ V_m $ & set of carrier- and phase-common nodes in mode $m \!\in\! M$ \\
      & \quad $(V_m = \cup_{p \in P} V_{mp})$ \\
      $ A_m $ & set of carrier- and phase-common directed arcs $m \!\in\! M$ \\
      & \quad $(A_m = \cup_{p \in P} A_{mp})$ \\
      $ V_p $ & set of nodes across modes for $p \in P$ \\
      & \quad $(V_p = \cup_{m \in M} V_{mp}; V_{p+1}^+ = V_{p}^-)$ \\
      $ A_p $ & set of directed arcs across modes for $p \in P$ \\
      & \quad $(A_p = \cup_{m \in M} A_{mp})$ \\
      $ V_k^+ $ & set of supply nodes in network $k \in M \cup P \cup \left(M \times P \right)$ \\
      & \quad $(V_k^+ = \{i \in V_k: b^k_i > 0 \})$\\
      $ V_k^- $ & set of demand nodes in network $k \in M \cup P \cup \left(M \times P \right)$ \\
      & \quad $(V_k^- = \{i \in V_k: b^k_i < 0 \})$\\
      $ N_k^+\!(\!i\!)\! $ & set of out-neighbors in network $k \in M \cup P \cup \left(M \times P \right)$ \\
      & \quad $(N_k^+(i) = \{j \in V_k: ij \in A_k\})$ \\
      $ N_k^-\!(\!i\!)\! $ & set of in-neighbors in network $k \in M \cup P \cup \left(M \times P \right)$ \\
      & \quad $(N_k^-(i) = \{j \in V_k: ji \in A_k\})$ \\
      \hline
    \end{tabular}
  \end{center}
\end{table}

\begin{table}[htbp!]
  \begin{center}
    \caption{Constants}
    \label{tab:constants}
    \scalebox{0.95}{
    \begin{tabular}{l|l|l|l}
      \textbf{Con.}\!\! & \!\!\textbf{Dom.}\!\!\! & \textbf{Unit} & \textbf{Description}\\
      \hline
      $ n_P$ & $\mathbb{Z}_+$ & \# & number of supply phases \\
      $ n_L$ & $\mathbb{Z}_+$ & \# & number of pieces in BPR approximation \\
      $ n_D^{mp} $ & $\mathbb{Z}_+$ & \# & defense supply capacity $m \!\!\in\!\! M, \!p \!\!\in\!\! P$ \\
      $ n_O^{mp} $ & $\mathbb{Z}_+$ & \# & open reserve capacity $m \!\!\in\!\! M, \!p \!\!\in\!\! P$ \\
      $ n_A^{mp} $ & $\mathbb{Z}_+$ & \# & attack supply capacity $m \!\!\in\!\! M, \!p \!\!\in\!\! P$ \\
      $c_{ij}^{cmp}$ & $\mathbb{R}_+$ & \$/(v/h) & flow cost $ij \!\!\in\!\! A_m, \!c \!\!\in\!\! C_{mp}, \!m \!\!\in\!\! M, \!p \!\!\in\!\! P$ \\
      $w_{ij}^{m}$ & $\mathbb{R}_+$ & \$/(\!(\!v/h)-\!h\!)\!\! & mode-time cost $ij \!\!\in\!\! A_m, \!m \!\!\in\!\! M$ \\      
      $u_{ij}^{cmp}$\! & $\mathbb{R}_+$ & v/h & flow capacity $ij \!\!\in\!\! A_{mp}, \!c \!\!\in\!\! C_{mp}, \!m \!\!\in\!\! M, \!p \!\!\in\!\! P$ \\
      $u_{ij}^m$ & $\mathbb{R}_+$ & v/h & carrier- and phase-agg cap $\!i\!j \!\!\in\!\! A_m\!, \!m \!\!\in\!\! M$ \\
    %   &&& \quad $\left(=\tfrac{h_{ij}^m \ell_{ij}^m}{\ell^{*mp}}\right)$ \\
      $b_i^{cmp}$\! & $\mathbb{R}$ & bbl/h & supply capacity $\!i \!\!\in\!\! V_{mp}\!, \!c \!\!\in\!\! C_{mp}\!, \!m \!\!\in\!\! M\!, \!p \!\!\in\!\! P$ \\
      &&& \quad $(\sgn(b_i^{cmp})\sgn(b_i^{mp}) \geq 0)$\\
      $b_i^{mp}$ & $\mathbb{R}$ & bbl/h & carrier-agg cap $i \!\!\in\!\! V_{mp}, \!m \!\!\in\!\! M\!, \!p \!\!\in\!\! P$ \\
      &&& \quad $(\sgn(b_i^{mp})\sgn(b_i^p) \geq 0)$\\
      $b_i^p$ & $\mathbb{R}$ & bbl/h & carrier- and mode-agg cap $i \!\!\in\!\! V_p,\! p \!\!\in\!\! P$ \\
      &&& \quad $(b_i^{p+1} = - b_i^p, \forall i \in V_p^-)$ \\
      $p_i^p$ & $\mathbb{R}_+$ & \$/(\!bbl/h\!)\! & carrier- and mode-agg penalty $\!\!i \!\!\in\!\! V_p\!, \!p \!\!\in\!\! P$ \\
      $v_{ij}^m$ & $\mathbb{R}_+$ & mi/h & max speed along $ij \in A_m, m \in M$ \\
      $\ell_{ij}^m$ & $\mathbb{R}_+$ & mi & length of arc $ij \in A_m, m \in M$ \\
      $h_{ij}^m$ & $\mathbb{Z}_+$ & \# & number of lanes $ij \in A_m, m \in M$ \\
      $q^m$ & $\mathbb{R}_+$ & h & max trip travel time $m \in M$ \\
      $\epsilon_{ij}^m$ & $\mathbb{R}_+$ & v/h & BPR set point width $ij \!\!\in\!\! A_m, \!m \!\!\in\!\! M$ \\ 
      &&& \quad $\left(\epsilon_{ij}^m = \tfrac{2u_{ij}^m}{n_L}\right)$ \\
    % just mode-time:
      $t_{ijr}^m$ & $\mathbb{R}_+$ & (v/h)-h & height $ij \!\in\! A_m, r \!\in\! L \cup \{0\}, m \!\in\! M$ \\
      &&& \quad $\left(\!T_{ij}^m(y)\! = \!\tfrac{\ell_{ij}^m}{v_{ij}^m}\!\left(\!1 \!+\!  0.15 \!\left(\!\tfrac{y}{u_{ij}^m} \!\right)^{\!4}\right)\!\right)$ \\
      &&& \quad $\left(t_{ijr}^m = r\epsilon_{ij}^m T_{ij}^m\left(r\epsilon_{ij}^m\right)\right)$ \\
      $\alpha_{ijr}^m\!$ & $\mathbb{R}_+$ & h & BPR slope $ij \in A_m, r \in L, m \in M$ \\
      &&& \quad $\left(\alpha_{ijr}^m  =\tfrac{t_{ijr}^m - t_{ijr-1}^m}{\epsilon_{ij}^m}\right)$ \\
      $\xi_{ijr}^m$ & $\mathbb{R}_-$ & h & BPR intercept $ij \!\in\! A_m, m \!\in\! M, r \!\in\! L$ \\
      &&& \quad $\left(\xi_{ijr}^m = t_{ijr}^m - \alpha_{ijr}^m r \epsilon_{ij}^m \right)$ \\
    %   &&& \quad $\left(\xi_{ijr}^m = -\left( (r-1)t_{ijr}^m + r t_{ijr-1}^m \right)\right)$ \\
      % should maybe be -(i y_i+1 - (1+i) y_i)?
      $\ell^{cmp}$ & $\mathbb{R}_+$ & mi/u & length of vehicle $c \!\in\! C_{mp}, m \!\in\! M\!, p \!\in\! P$ \\
      $\ell^{*m}$ & $\mathbb{R}_+$ & mi/v & length of standard vehicle in mode $m \in M$ \\
      $\rho^{cmp}$ & $\mathbb{R}_+$ & bbl/u & demand per vehicle $c \!\!\in\!\! C_{mp}, m \!\!\in\!\! M\!, p \!\!\in\!\! P$ \\
      $\gamma^{cmp}$ & $\mathbb{R}_+$ & v/bbl & supply unit conversion $c \!\!\in\!\! C_{mp}, m \!\!\in\!\! M\!, p \!\!\in\!\! P$ \\
    %   &&& \quad $\left(=\tfrac{\ell^{cmp}}{\ell^{*mp} \rho^{cmp}}\right)$ \\
      $ \nu_i^{p} $ & $\mathbb{Z}_+$ & \# & number of pumps at $i \in V_p, p \in P$ \\
      $ \psi_i^{p} $ & $\mathbb{R}_+$ & bbl/h & service rate per pump $i \in V_p, p \in P$ \\
      \hline
    \end{tabular}
    }
  \end{center}
\end{table}

\begin{table}[htbp!]
  \begin{center}
    \caption{Variables}
    \label{tab:variables}
    \begin{tabular}{l|l|l|l}
      \textbf{Var.}\! & \!\!\textbf{Dom.}\!\!\! & \textbf{Unit} & \textbf{Description}\\
      \hline
      % defender
      $d_i^{mp}$ & $\{\!0,\! 1\!\}$\! & \# & defend supply decision $\!i \!\!\in\!\! S_{mp}\!, \!m \!\!\in\!\! M\!\!, \!p \!\!\in\!\! P$ \\
      $o_i^{mp}$ & $\{\!0,\! 1\!\}$\! & \# & open reserve decision $i \!\!\in\!\! R_{mp}, \!m \!\!\in\!\! M\!, \!p \!\!\in\!\! P$ \\
      % attacker
      $a_i^{mp}$ & $\{\!0,\! 1\!\}$\! & \# & attack supply decision $i \!\!\in\!\! S_{mp}, \!m \!\!\in\!\! M\!, \!p \!\!\in\!\! P$ \\
      $f_{ij}^{cmp}$\! & $\mathbb{R}_+$ & v/h & flow for $ij \!\!\in\!\! A_{mp}, \!c \!\!\in\!\! C_{mp}, \!m \!\!\in\!\! M\!, \!p \!\!\in\!\! P$ \\
    %   $f_{ij}^{mp}$ & $\mathbb{R}_+$ & v/h & carrier-aggregated flow $\!i\!j \!\!\in\!\! A_m\!, \!m \!\!\in\!\! M\!, \!p \!\!\in\!\! P$ \\
      $f_{ij}^m$ & $\mathbb{R}_+$ & v/h & carrier- and phase-agg flow $ij \!\!\in\!\! A_m, \!m \!\!\in\!\! M$ \\
      $x_i^{cmp}$\! & $\mathbb{R}_+$ & bbl/h & supply for $i \!\!\in\!\! V_{mp}, \!c \!\!\in\!\! C_{mp}, \!m \!\!\in\!\! M, \!p \!\!\in\!\! P$ \\
      $x_i^{mp}$ & $\mathbb{R}_+$ & bbl/h & carrier-aggregated sup $i \!\!\in\!\! V_{mp}, \!m \!\!\in\!\! M, p \!\!\in\!\! P$ \\
      $x_i^p$ & $\mathbb{R}_+$ & bbl/h & carrier- and mode-agg sup $i \!\!\in\!\! V_p,\! p \!\!\in\!\! P$ \\
    %   $s_i^{cmp}$\! & $\mathbb{R}_+$ & bbl/h & supply slack $i \!\!\in\!\! V_m, \!c \!\!\in\!\! C_p, \!m \!\!\in\!\! M, \!p \!\!\in\!\! P$ \\
      $s_i^{mp}$ & $\mathbb{R}_+$ & bbl/h & carrier-agg slack $i \!\!\in\!\! V_{mp}, \!m \!\!\in\!\! M, \!p \!\!\in\!\! P$ \\
    %   $s_i^p$ & $\mathbb{R}_+$ & bbl/h & carrier- and mode-agg slack $i \!\!\in\!\! V_p, \!p \!\!\in\!\! P$ \\
    %   $t_{ij}^{m}$ & $\mathbb{R}_+$ & aggregate time $ij \!\!\in\!\! A_m, \!m \!\!\in\!\! M$ \\
      $g_{ij}^{m}$ & $\mathbb{R}_+$ & (\!v/h\!)-h\!\! & aggregate mode-time $ij \!\!\in\!\! A_m, \!m \!\!\in\!\! M$ \\
    % duals
    $\phi_i^{cmp}$\! & $\mathbb{R}$ & - & dual $i \!\!\in\!\! V_{mp}, \!c \!\!\in\!\! C_{mp}, \!m \!\!\in\!\! M, \!p \!\!\in\!\! P$ \\      
    $\kappa_{ij}^{cmp}$\! & $\mathbb{R}$ & - & dual $ij \!\!\in\!\! A_{mp}, \!c \!\!\in\!\! C_{mp}, \!m \!\!\in\!\! M, \!p \!\!\in\!\! P$ \\      
    $\beta_i^{cmp}$\! & $\mathbb{R}_+$ & - & dual $i \!\!\in\!\! V_{mp}, \!c \!\!\in\!\! C_{mp}, \!m \!\!\in\!\! M, \!p \!\!\in\!\! P$ \\  $\mu_{ij}^{cmp}$\! & $\mathbb{R}_+$ & - & dual $ij \!\!\in\!\! A_{mp}, \!c \!\!\in\!\! C_{mp}, \!m \!\!\in\!\! M, \!p \!\!\in\!\! P$ \\    
    $\sigma_i^{mp}$\! & $\mathbb{R}$ & - & dual $i \!\!\in\!\! V_{mp}, \!m \!\!\in\!\! M, \!p \!\!\in\!\! P$ \\
    $\delta_i^{mp}$\! & $\mathbb{R}_+$ & - & dual $i \!\!\in\!\! S_{mp}, \!m \!\!\in\!\! M, \!p \!\!\in\!\! P$ \\
    $\omega_i^{mp}$\! & $\mathbb{R}$ & - & dual $i \!\!\in\!\! R_{mp}, \!m \!\!\in\!\! M, \!p \!\!\in\!\! P$ \\
    $\beta_i^{mp}$\! & $\mathbb{R}$ & - & dual $i \!\!\in\!\! V_{mp}\!\setminus\! R_{mp}, \!m \!\!\in\!\! M, \!p \!\!\in\!\! P$ \\
    $\sigma_i^{p}$\! & $\mathbb{R}$ & - & dual $i \!\!\in\!\! V_p, \!p \!\!\in\!\! P$ \\
    $\beta_i^{p}$\! & $\mathbb{R}_+$ & - & dual $i \!\!\in\!\! V_p, \!p \!\!\in\!\! P$ \\
    $\kappa_{ij}^{m}$\! & $\mathbb{R}$ & - & dual $ij \!\!\in\!\! A_m, \!m \!\!\in\!\! M$ \\
    $\mu_{ij}^{m}$\! & $\mathbb{R}_+$ & - & dual $ij \!\!\in\!\! A_m, \!m \!\!\in\!\! M$ \\
    $\tau_{ijr}^{m}$\! & $\mathbb{R}_+$ & - & dual $ij \!\!\in\!\! A_m, \!m \!\!\in\!\! M, r \in L$ \\
    $\upsilon_i^{p}$\! & $\mathbb{R}_+$ & - & dual $i \!\!\in\!\! V_p, \!p \!\!\in\!\! P$ \\
    $\theta^{stm}$\! & $\mathbb{R}$ & - & dual $st \in C_{mn_P}, m \in M$ \\
      \hline
    \end{tabular}
  \end{center}
\end{table}

\begin{table}[htbp!]
  \begin{center}
    \caption{Parameter Values}
    \label{tab:params}
    \begin{tabular}{l|l}
      \textbf{Param.} & \textbf{Value}\\
      \hline
      % sets
      $ n_P$ & 3 \\
      $ n_L$ & 4 \\
      $ M $ & $\{\mathcal{P}, \mathcal{T}\}$ \\
      $ C_{mp} $ & $C_{m1} = \{m\}$, $ C_{m2} = V_{m2}^- \!\times\! V_{m2}^+, C_{m3} = V_{m3}^- \!\times\! V_{m3}^+$ \\
    %   $ V_m $ & ? \\
    %   $ A_m $ & ? \\
    %   $ R_{mp} $ & ? \\
      $S_{mp}$ & $S_{mp} \subseteq V_{mp}^+$ \\
      $R_{mp}$ & $R_{mp} \subseteq V_{mp}^+ \setminus S_{mp}$ \\
      % constants
    %   $ n_D^{mp} $ & $ 0 \leq n_D^{mp} \leq ?, \forall ?$ \\
    %   $ n_O^{mp} $ & $ 0 \leq n_O^{mp} \leq ?, \forall ?$ \\
    %   $ n_A^{mp} $ & $ 0 \leq n_A^{mp} \leq ?, \forall ?$ \\
      % units all the same: bbl
      $c_{ij}^{cmp}$ & $c_{ij}^{cmp} = c_{ij}^{c'mp}, \forall c, c' \in C_{mp}, c_{ij}^{cm2} = c_{ij}^{cm3} = \tfrac{q}{2}$ \\
      %$u_{ij}^m$ & ? \\
    %   $u_{ij}^{mp}$ & $u_{ij}^{mp} = u_{ij}^m, \forall m \in M, p \in P$ \\
      $\gamma^{cmp}$ & $\gamma^{cmp} = \tfrac{\ell^{cmp}}{\ell^{*m}\rho^{cmp}}$ \\
      $u_{ij}^m$ & $u_{ij}^m = \tfrac{h_{ij}^m v_{ij}^m}{\ell^{*m}}$ \\
      $u_{ij}^{cmp}$ & $u_{ij}^{cmp} = u_{ij}^{m}$ \\
      & (separate truck from customer $u$ at gas station) \\
      $b_i^p$ & $b_i^3 = -b_i^2, \forall i \in V_2^+$ \\
      $b_i^{mp}$ & $b_i^{mp} = b_i^p, i \in V_{mp}$ \\
      $b_i^{cmp}$ & $b_i^{cm1} = b_i^{m1}$ \\
      & $b_i^{stmp} = b_i^{mp} \mathbb{I}_{i \!\in\! \{\!s, t\!\}} \!, st \!\in\! C_{mp}, \!p \in \!\{2, 3\}$ \\
    %   $p_i^{cmp}$ & $p_i^{cmp} = 0, c \in C_p, m \in M, p \in P$ \\
      $p_i^{mp}$ & $p_i^{m2} = p_{i}^{m3}$ \\
    %   $p_i^p$ & $p_i^2 = p_i^3$ \\
      \hline
    \end{tabular}
  \end{center}
\end{table}
%
% (TODO: first describe the generic model in words with symbol definitions that show up in our tables)

% (TODO: then reference the last table that has our problem-specific constraints/choices.)

The particular operator model we incorporate is a generalized variant of the fuel and transportation combo-model in Kuc's work~\cite{kuc2020} (pp.~69-71), which was developed based on the interdependent network flow problem introduced in~\cite{ahangar2020modeling} by Ahangar et al.  Generically, we allow for multiple modes of transportation ($M$) and multiple phases of supply ($P$).  For each mode ($m \in M$) and phase ($p \in P$), we let $G_{mp} = (V_{mp}, A_{mp})$ denote a directed graph upon which supply will be routed from supply to demand nodes, which are respectively defined by whether a node's capacity ($b_{i_{mp}}$) is positive or negative.  Here, corresponding networks are comprised of distinct arcs ($A_m, \forall m \in M$) and potentially overlapping nodes ($V_m$). Modes are meant to help enforce the requirement that shared arcs have no impact across systems, while phases are meant to help distinguish successive dependencies in the supply chain; however, vehicles in all modes and phases may differ across or even within a mode-supply pair.  In particular, we allow carrier types ($C_{mp}, \forall m \in M, p \in P$) to handle the routing of supplies, where each carrier type ($c \in C_{mp}$) has a vehicle length ($\ell^{cmp}$) and capacity ($\rho^{cmp}$). 
Second, we allow for multiple supply phases ($P = \{1, 2, \ldots n_P\}$) during which networks are kept static ($V_{mp} = V_{m}, A_{mp} = A_p, \forall m \in M, p \in P$), supply and demands are common across modes ($V_{mp}^+ \subseteq V_p^+, V_{mp}^- \subseteq V_p^-, \forall m \in M, p \in P$, with $V_k^+ = \{k \in V_k: b_i^k > 0\}$ and $V_k^- = \{k \in V_k: b_i^k < 0\}$), and each successive phase's supply equals the demand filled prior ($V_{p+1}^+ = V_p^-, \forall p \in P \setminus \{n_P\}$).  Third, upon each mode and phase, we supply fuel demands through various carrier types ($C_{mp}, \forall m \in M, p \in P$).  

\begin{figure*}
    \centering
    \includegraphics[width = \linewidth]{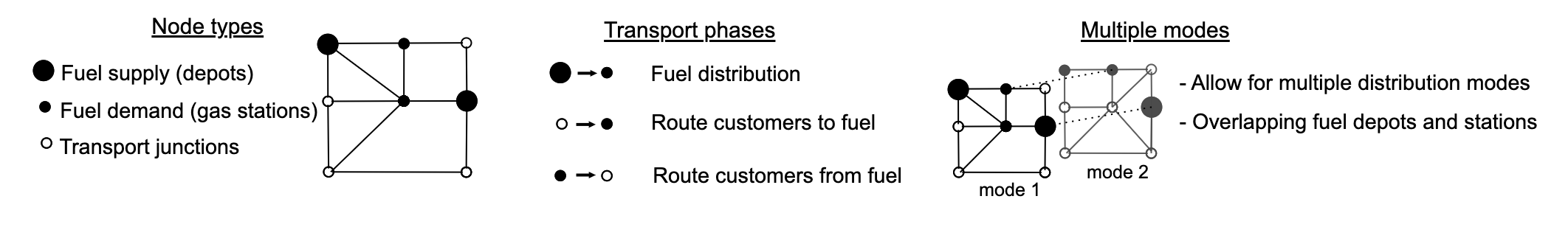}
    \caption{Schematic of three phases within a single transportation mode for fuel and transportation interdependent networks on the same topology, which we interpret as a road network. Note: our mathematical framework accommodates multiple fuel transportation modes (e.g, truck and rail), but in our initial experiments we focus on a single mode of transportation (i.e., routing by truck).}
    \label{fig:phases}
\end{figure*}

Our specific instance of this model aims to deliver fuel to gas stations during the first phase, through tanker trucks where carriers are of a single type ($C_{m1} = \{m\}, \forall m \in M$), while the second and third phases consist of customers being routed to and from such gas stations, respectively, and where carrier types correspond to customer distribution $(C_{mp} = V_{mp}^- \times V_{mp}^+, \forall m \in M, p \in P$). See Figure~\ref{fig:phases} for a snapshot within a single mode across the three phases, as well as an illustration of two modes. We embed such a model in the DAD framework, allowing attackers to choose to fully disrupt a subset of supply nodes within modes and phases ($S_{mp} \subseteq V_{mp}^+$), while the defender can mitigate such attacks, or open reserve supplies to augment supply shortages ($R_{mp} \subseteq V_{mp}^+ \setminus R_{mp}$).  Solving such a formulation aims to find the best way to meet demand during each phase (through each participating mode) by imposing penalty costs on unmet demand, while simultaneously balancing overall travel time across travelers in each mode.  Since our routing problem is optimized implicitly over a time horizon, our core units are in average rates (bbl/h) rather than instantaneous quantities (bbl), a choice that also allows the encoding of supply and demand delays together with traveler traffic congestion.
% For the defender-attacker-defender problem of ? problem we have three actors making decisions in sequence:
% %
% \begin{itemize}
% 	\item First, a defender chooses a ? defend, where defending means an attack on it is rendered completely inconsequential
% 	\item Second, an attacker chooses a ?, where attacking means ? (if not already defended)
% 	\item Finally, an operator attempts to operate the ?
% \end{itemize}

This can initially be modeled as \eqref{eqn:dad} where we define the defender's decision space, $W$, to be the set of vectors $d, o$ %\in \mathbb{R}$ 
satisfying the following constraints:
\begin{align}
    \textstyle\sum_{i \in S_{mp}} d_i^{mp} &\leq n_D^{mp} && \forall m \in M, p \in P \label{eqn:D1} \\
    \textstyle\sum_{i \in R_{mp}} o_i^{mp} &\leq n_O^{mp} && \forall m \in M, p \in P \label{eqn:D2} \\
    d_{i}^{mp} &\in \{0, 1\} && \forall i \in S_{mp}, m \in M, p \in P \label{eqn:D3}\\
    o_{i}^{mp} &\in \{0, 1\} && \forall i \in R_{mp}, m \in M, p \in P \label{eqn:D4}
\end{align}
The above constraints can be interpreted to mean the defender is limited to defending \eqref{eqn:D1} and opening \eqref{eqn:D2} at most $n_D^{mp}$ and $n_O^{mp}$ supply nodes, respectively, in any mode or phase, and defense decisions are all or nothing \eqref{eqn:D3}-\eqref{eqn:D4}.
Now we define the attacker's decision space, $X$, (which is structurally similar to $W$), to be the set of vectors $a$ %\in \mathbb{R}$ 
satisfying the following constraints:
\begin{align}
    \textstyle\sum_{i \in S_{mp}} a_i^{mp} &\leq n_A^{mp} && \forall m \in M, p \in P \label{eqn:A1} \\
    a_{i}^{mp} &\in \{0, 1\} && \forall i \!\in\! S_{mp}, m \!\in\! M, p \!\in\! P \label{eqn:A2}
\end{align}
These constraints can be interpreted to mean the attacker is limited to attacking \eqref{eqn:A1} at most $n_A^{mp}$ supply nodes in any mode or phase, and attack decisions are all or nothing \eqref{eqn:A2}.

% (TODO: Reword for our application) 
Note that rather than coupling attacker and defender constraints to encode protected supply nodes as not attackable, we decouple by encoding the effects thereof.  In other words, we allow the attacker to overlap chosen nodes with the defender, but, as we will see, the effects are nullified.
Finally, we define the operator's decisions space, $Y(w, x)$ for any $w \in W$, $x \in X$, to be the set of vectors $f, g, x, y$ %\in \mathbb{R}$ 
satisfying the following sets of constraints:
\begin{align}
  % link fuel flows: f and x
  \textstyle\sum_{j \!\in\! N_{mp}^+\!(i)}\hat{f}_{ij}^{cmp} \!-\!\! \textstyle\sum_{j \!\in\! N_{mp}^-\!(i)} \hat{f}_{ji}^{cmp} &= \sgn\!\left(\!b_i^{cmp}\!\right)\! x_i^{cmp} \\
  &\!\!\!\!\!\!\!\!\!\forall i \!\!\in\!\! V_{mp}, \!c \!\!\in\!\! C_{mp}, \!m \!\!\in\!\! M\!, \!p \!\!\in\!\! P \nonumber
\end{align}
\begin{align}
  % link fuel flow with common vehicle flow
  f_{ij}^{cmp} &= \gamma^{cmp}\hat{f}_{ij}^{cmp} && \!\forall ij \!\!\in\!\! A_{mp}, \!c \!\!\in\!\! C_{mp}, \!m \!\!\in\!\! M\!, \!p \!\!\in\!\! P \\
% \end{align}
% \begin{align}
  % over C x M x P: arc/node flow, slack, bounds, and link to next aggregate
%   x_i^{cmp} \!\!+\! s_i^{cmp} &= \left|b_{cmp}\right| && \!\forall i \!\!\in\!\! V_m, \!c \!\!\in\!\! C_p, \!m \!\!\in\!\! M\!, \!p \!\!\in\!\! P \\
  x_i^{cmp} &\leq \left|b_i^{cmp}\right| && \!\forall i \!\!\in\!\! V_{mp}, \!c \!\!\in\!\! C_{mp}, \!m \!\!\in\!\! M\!, \!p \!\!\in\!\! P \\
%   f_{ij}^{cmp} &\leq u_{ij}^{cmp} && \!\forall ij \!\!\in\!\! A_m, \!c \!\!\in\!\! C_{mp}, \!m \!\!\in\!\! M\!, \!p \!\!\in\!\! P\\
  f_{ij}^{cmp} &\leq 2u_{ij}^{cmp} && \!\forall ij \!\!\in\!\! A_{mp}, \!c \!\!\in\!\! C_{mp}, \!m \!\!\in\!\! M\!, \!p \!\!\in\!\! P\\
%   f_{ij}^{cmp} &\geq 0 && \!\forall ij \!\!\in\!\! A_m, \!c \!\!\in\!\! C_{mp}, \!m \!\!\in\!\! M\!, \!p \!\!\in\!\! P \\
  f_{ij}^{cmp} &\geq 0 && \!\forall ij \!\!\in\!\! A_{mp}, \!c \!\!\in\!\! C_{mp}, \!m \!\!\in\!\! M\!, \!p \!\!\in\!\! P \\
  x_i^{cmp} &\geq 0 && \!\forall i \!\!\in\!\! V_{mp}, \!c \!\!\in\!\! C_{mp}, \!m \!\!\in\!\! M\!, \!p \!\!\in\!\! P
%   s_i^{cmp} &\geq 0 && \!\forall i \!\!\in\!\! V_m, \!c \!\!\in\!\! C_{mp}, \!m \!\!\in\!\! M\!, \!p \!\!\in\!\! P
\end{align}
% \begin{align}
%   % over M x P: arc/node flow, slack, bounds, and link to next aggregate
%   x_i^{mp} \!+\! s_i^{mp} &= \left|b_{mp}\right| && \forall i \!\!\in\!\! V_m, \!m \!\!\in\!\! M\!, \!p \!\!\in\!\! P \\
%   f_{ij}^{mp} &\leq u_{ij}^{mp} && \forall ij \!\!\in\!\! A_m, \!m \!\!\in\!\! M\!, \!p \!\!\in\!\! P\\
% %   x_i^{cmp} &\leq \left|b_{cmp}\right| && \forall i \!\!\in\!\! V_m, \!c \!\!\in\!\! C_p, \!m \!\!\in\!\! M, \!p \!\!\in\!\! P \\  
%   f_{ij}^{mp} &= \textstyle\sum_{c \in C_p} f_{ij}^{cmp} && \forall ij \!\!\in\!\! A_m, \!m \!\!\in\!\! M\!, \!p \!\!\in\!\! P\\
%   x_i^{mp} &= \textstyle\sum_{c \in C_{mp}} x_i^{cmp} && \forall i \!\!\in\!\! V_m, \!m \!\!\in\!\! M\!, \!p \!\!\in\!\! P
%   s_i^{mp} &\geq 0 && \forall i \!\!\in\!\! V_m, \!m \!\!\in\!\! M\!, \!p \!\!\in\!\! P
% \end{align}
\begin{align}
    x_i^{mp} &= \textstyle\sum_{c \in C_{mp}} x_i^{cmp} && \!\forall i \!\!\in\!\! V_{mp}, \!m \!\!\in\!\! M\!, \!p \!\!\in\!\! P \\
  % over M x P: link node flow bounds with defense, reserve, and attack effects
%   x_i^{mp}\!+\!s_i^{mp} \!&=\! \left|b_i^{mp}\right|\! \left(\!1 \!-\! \left(\!1 \right.\right.&& \!\!\!\!\!\!\!\!\!\!\!\!\!\!\!\!\!\left.\left.\!-\! d_i^{mp}\!\right)\!a_i^{mp}\right) \\
%   &&& \!\forall i \!\!\in\!\! S_{mp}, \!m \!\!\in\!\! M\!, \!p \!\!\in\!\! P \nonumber\\
  x_i^{mp} \!&\leq\! \left|b_i^{mp}\right|\! \left(\!1 \!-\! \left(\!1 \!-\! d_i^{mp}\!\right)\!a_i^{mp}\right) && \!\forall i \!\!\in\!\! S_{mp}, \!m \!\!\in\!\! M\!, \!p \!\!\in\!\! P \\
  x_i^{mp}\!+\!s_i^{mp} \!&=\! \left|b_i^{mp}\right| \! o_i^{mp} && \!\forall i \!\!\in\!\! R_{mp}\!, \!m \!\!\in\!\! M\!, \!p \!\!\in\!\! P    
\end{align}
\begin{align}
    % x_i^{mp}\!+\!s_i^{mp} \!&=\! \left|b_i^{mp}\right| && \!\forall i \!\!\in\!\! V_{mp}\!\!\!\setminus\!\!\left(\!S_{mp} \!\!\cup\!\! R_{mp}\!\right)\!\!, \!m \!\!\in\!\! M\!\!, \!p \!\!\in\!\! P \\
    x_i^{mp}\!+\!s_i^{mp} \!&=\! \left|b_i^{mp}\right| && \!\forall i \!\!\in\!\! V_{mp}\!\!\!\setminus\!\! R_{mp}\!, \!m \!\!\in\!\! M\!\!, \!p \!\!\in\!\! P \\
    s_i^{mp} &\geq 0 && \!\forall i \in V_{mp}, m \in M, p \in P 
\end{align}
\begin{figure*}[h!]
    \centering
    \includegraphics[width=0.80\linewidth]{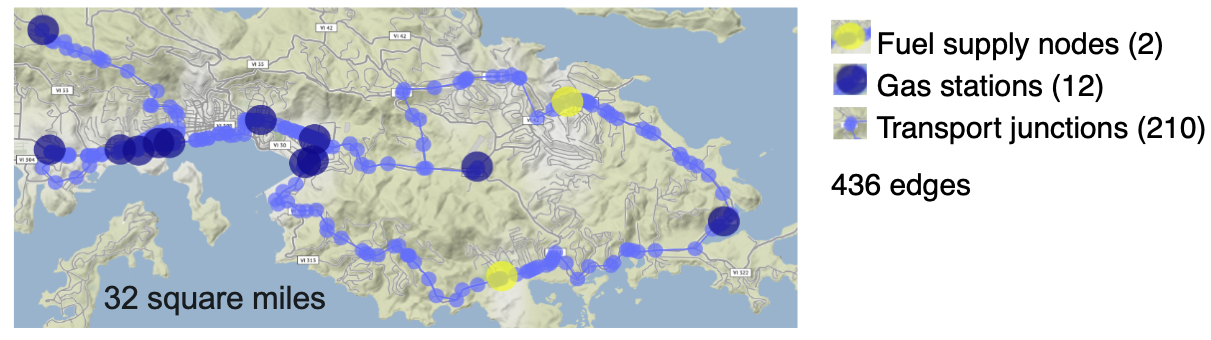}
    \caption{Interdependent fuel and transportation networks at St. Thomas, USVI.}
    \label{fig:usvi}
\end{figure*}
\begin{align}
  % over P: node flow, slack, bounds
  x_i^{p} &= \textstyle\sum_{m \in M: i \in V_{mp}} x_i^{mp} && \forall i \!\!\in\!\! V_p, \!p \!\!\in\!\! P \\
  x_i^{p} &\leq \left|b^p_i\right| && \forall i \!\!\in\!\! V_p, \!p \!\!\in\!\! P \\
%   x_i^{cmp} &\leq \left|b_{cmp}\right| && \forall i \!\!\in\!\! V_m, \!c \!\!\in\!\! C_p, \!m \!\!\in\!\! M, \!p \!\!\in\!\! P \\ 
%   s_i^{p} &\geq 0 && \forall i \!\!\in\!\! V_p, \!p \!\!\in\!\! P \\
% \end{align}
% \begin{align}
  % over M: arc flow, bounds
  f_{ij}^{m} &= \sum_{p \in P} \sum_{c \in C_{mp}} f_{ij}^{cmp} && \forall ij \!\!\in\!\! A_m, \!m \!\!\in\!\! M \\
  f_{ij}^{m} &\leq 2u_{ij}^{m} && \forall ij \!\!\in\!\! A_m, \!m \!\!\in\!\! M \\
% \end{align}
% \begin{align}
  % over M: link arc flow with convex piecewise linear approx of BPR (v * S(v)) to mode-hours (congestion)
%   t_{ij}^m &\geq \tfrac{\alpha_{ijr}^m}{\gamma^m} f_{ij}^m + \beta_{ijr} && \forall ij \!\!\in\!\! A_m, \!m \!\!\in\!\! M \\
  % TODO: figure out if linear transformation exists between two (probably not)
  g_{ij}^m &\geq \alpha_{ijr}^m f_{ij}^m + \xi_{ijr}^m && \forall ij \!\!\in\!\! A_m, \!m \!\!\in\!\! M, \!r \!\!\in\!\! L \\
% \end{align}
% \begin{align}
  % over successive P: demand met becomes next supply 
  % TODO: (does this imply the symmetry we want x^st = x^ts??)
  x_i^p &\geq x_i^{p+1} && \forall i \!\!\in\!\! V_p^-, \!p \!\!\in\!\! P \!\setminus\!\!\{\!n_P\!\} \\
  % specific to our use case: tie last two phases' supplies since they are leave/return routes
  % may be overly redundant (replace i with s/t?)
  x_s^{stmn_P} &= x_t^{tsm,n_P-1} && \forall st \!\!\in \!\! C_{mn_P},\! m \!\!\in \!\! M
\end{align}
with objective function \eqref{eqn:longobj} taking the form:
\begin{align}\label{eqn:longobj}
    \sum_{m \in M} \sum_{p\in P} \sum_{c \in C_{mp}} \sum_{ij \in A_{mp}} c_{ij}^{cmp} f_{ij}^{cmp} \; & + \nonumber \\
    \sum_{m \in M} \sum_{p \in P} \sum_{i \in V_{mp}} p_i^{mp} s_i^{mp} \; & + \\
    \sum_{m \in M} \sum_{ij \in A_m} w_{ij}^m g_{ij}^m \; & \nonumber
\end{align}
%
% These constraints can be interpreted to mean the operator is ?.

The dual problem is:
\begin{align}
    \phi_i^{cmp} \!\!-\!\! \phi_j^{cmp} \!\!-\!\! \gamma^{cmp} \!\kappa_{ij}^{cmp} & \leq 0 \\
    & \forall ij \!\!\in\!\! A_{mp}, \!c \!\!\in\!\! C_{mp}, \!m \!\!\in\!\! M\!, \!p \!\!\in\!\! P \nonumber \\
    \kappa_{ij}^{cmp} - \kappa_{ij}^m - \mu_{ij}^{cmp} & \leq c_{ij}^{cmp} \\
    & \forall ij \!\!\in\!\! A_{mp}, \!c \!\!\in\!\! C_{mp}, \!m \!\!\in\!\! M\!, \!p \!\!\in\!\! P \nonumber
\end{align}
\begin{align}
    \kappa_{ij}^m \!-\! \mu_{ij}^m \!-\! \sum_{r \in L} \alpha_{ijr}^m \tau_{ijr}^m & \leq 0 && \forall ij \!\!\in\!\! A_m, m \!\!\in\!\! M 
\end{align}
\begin{align}
    - & sgn\left(b_i^{cmp}\right) \phi_i^{cmp} - \beta_i^{cmp} - \sigma_i^{mp} \nonumber \\
    + & \theta^{stm} \mathbb{I}_{p=n_P, c=st, i=s} - \theta^{tsm} \mathbb{I}_{p=n_P-1, c=ts, i=t} \leq 0 \\
    & \quad \forall i \!\!\in\!\! V_{mp}, \!c \!\!\in\!\! C_{mp}, \!m \!\!\in\!\! M\!, \!p \!\!\in\!\! P \nonumber 
\end{align}
\begin{align}
    & \sigma_i^{mp} - \sigma_i^p - \delta_i^{mp} \mathbb{I}_{i \in S_{mp}} + \omega_i^{mp} \mathbb{I}_{i \in R_{mp}} \nonumber \\ 
    + & \beta_i^{mp} \mathbb{I}_{i \in V_{mp} \setminus R_{mp}} \leq 0 \\
    & \quad \forall i \!\!\in\!\! V_{mp}, m \!\!\in\!\! M, p \!\!\in\!\! P \nonumber 
\end{align}
\begin{align}
    \sigma_i^p \!-\! \beta_i^p \!+\! \upsilon_i^p \mathbb{I}_{p \neq n_P, i \in V_p^-} \!-\! \upsilon_i^{p-1} \mathbb{I}_{p \neq 1, i \in V_{p-1}^-} \leq 0 && \forall i \!\!\in\!\! V_p\!, \!p \!\!\in\!\! P 
\end{align}
\begin{align}
    % & \delta_i^{mp} \mathbb{I}_{i \in S_{mp}} \!+\! \omega_i^{mp} \mathbb{I}_{i \in R_{mp}} \\
    & \omega_i^{mp} \mathbb{I}_{i \in R_{mp}} + \beta_i^{mp} \mathbb{I}_{i \in V_{mp} \!\setminus\! R_{mp}} \leq p_i^{mp} && \forall i \!\! \in \!\! V_{mp}\!, m \!\! \in \!\! M\!, p \!\!\in \!\! P \nonumber
\end{align}
\begin{align}
    \sum_{r \in L} \tau_{ijr}^m \leq w_{ij}^m && \forall ij \in A_m, m \in M
\end{align}
with objective
\begin{align}
    - \sum_{m \in M} \sum_{p \in P} \sum_{c \in C_{mp}} \sum_{i \in V_{mp}} |b_i^{cmp}| \beta_i^{cmp} \; & \nonumber \\
    - \sum_{m \in M} \sum_{p \in P} \sum_{c \in C_{mp}} \sum_{ij \in A_{mp}} 2u_{ij}^{cmp} \mu_{ij}^{cmp} \; & \nonumber \\
    - \sum_{m \in M} \sum_{p \in P} \sum_{i \in S_{mp}} |b_i^{mp}| \overline{\delta}_i^{mp} \; & \nonumber \\
    + \sum_{m \in M} \sum_{p \in P} \sum_{i \in R_{mp}} |b_i^{mp}| o_i^{mp} \omega_i^{mp} \; & \\
    + \sum_{m \in M} \sum_{p \in P} \sum_{i \in V_{mp} \setminus R_{mp}} |b_i^{mp}| \beta_i^{mp} \; & \nonumber \\
    - \sum_{p \in P} \sum_{i \in V_p} |b_i^p| \beta_i^p \nonumber\\
    - \sum_{m \in M} \sum_{ij \in A_m} 2u_{ij}^m \mu_{ij}^m \; & \nonumber \\
    + \sum_{m \in M} \sum_{ij \in A_m} \sum_{r \in L} \xi_{ijr}^m \tau_{ijr}^m \; & \nonumber
\end{align}
where $\overline{\delta}_i^{mp} \in \mathbb{R}_+$, and
\begin{align}
    \overline{\delta}_i^{mp} &= \left(1 - (1 - d_i^{mp}) a_i^{mp} \right) \delta_i^{mp} && \forall i \!\! \in \!\! S_{mp}\!, \!m \!\! \in \!\! M\!, \!p \!\!\in \!\! P
\end{align}
This constraint is bilinear, but can be linearized with a well-known big-M method as follows:
\begin{align}
    % \overline{\delta}_i^{mp} & \!\!\geq\!\! -\!M \!\left(\!1 \!\!-\!\! (\!1 \!\!-\!\! d_i^{mp}) a_i^{mp} \!\right) \!\forall i \!\! \in \!\! S_{mp}\!, \!m \!\! \in \!\! M\!, \!p \!\!\in \!\! P \\
    \overline{\delta}_i^{mp} & \!\!\geq 0 && \forall i \!\! \in \!\! S_{mp}\!, \!m \!\! \in \!\! M\!, \!p \!\!\in \!\! P \\
    \overline{\delta}_i^{mp} & \!\leq \!M \!\left(\!1 \!\!-\!\! (\!1 \!\!-\!\! d_i^{mp}) a_i^{mp} \!\right) && \forall i \!\! \in \!\! S_{mp}\!, \!m \!\! \in \!\! M\!, \!p \!\!\in \!\! P\\
    % \delta_i^{mp} \!\!-\!\! \overline{\delta}_i^{mp} & \!\geq\! -\!M \!\left(\!1 \!\!-\!\! d_i^{mp}\right) a_i^{mp} \quad\forall i \!\! \in \!\! S_{mp}\!, \!m \!\! \in \!\! M\!, \!p \!\!\in \!\! P\\
    \delta_i^{mp} \!\!-\!\! \overline{\delta}_i^{mp} & \!\geq\! 0 && \forall i \!\! \in \!\! S_{mp}\!, \!m \!\! \in \!\! M\!, \!p \!\!\in \!\! P\\
    \delta_i^{mp} \!\!-\!\! \overline{\delta}_i^{mp} & \!\leq\! M \!\left(\!1 \!\!-\!\! d_i^{mp}\right) a_i^{mp} && \forall i \!\! \in \!\! S_{mp}\!, \!m \!\! \in \!\! M\!, \!p \!\!\in \!\! P    
\end{align}
The value of $M$ or $-M$ may be tightened if bounds can be computed in advance, i.e. if $-M \leq \ell \leq \delta_i^{mp} \leq u \leq M$ for some $\ell, u$ with $\ell \leq 0 \leq u$. (This may be the case for the upper bound, by (37) $\delta \leq p$, and if S and R are disjoint, then all deltas are bounded above by this).
\begin{figure}
    \centering
    \includegraphics[width=0.99\linewidth]{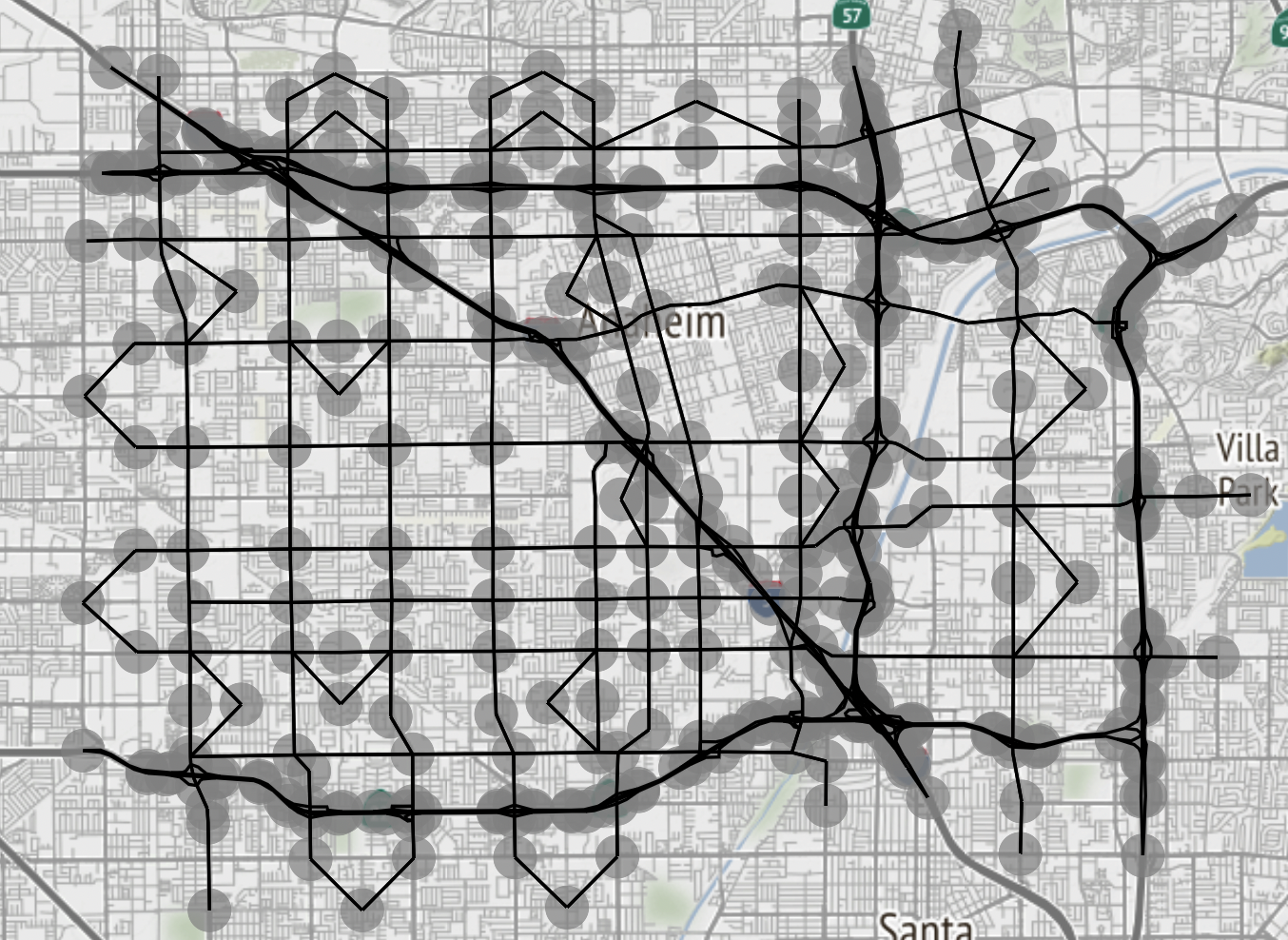}
    \caption{The Anaheim road network with 416 nodes and 914 edges.}
    \label{fig:anaheim}
\end{figure}

% (TODO: mention if/how we speed things up with ML active set learning on attacker/defender decisions?)

\section{Numerical Case Studies}
To implement our DAD model, we perform numerical experiments with hypothetical use cases involving defense, reserve, and attack budget scenarios based on fictitious, yet realistic data generated by the Naval Postgraduate School for the U.S. Virgin Island (USVI) of St. Thomas~\cite{kuc2020}, as well as on the widely-used realistic Anaheim transportation network of~\cite{stabler2018transportation}. Our experiments with the St. Thomas and Anaheim data illustrate the applicability of our framework on a metropolitan scale while providing practical insight for critical infrastructure defense planning, as well as its ability to generalize to various settings. To further investigate scalability and generalizability, we ran our model on multiple networks generated as a) samples from graphs on $N$ nodes with power law degree distributions~\cite{barabasi1999emergence} b) samples drawn from graphs on $N$ nodes with an exponential degree distribution~\cite{zhang2017backbone}, and c) samples from the Grid network with Random Edges and Regional Costs (GREREC) random road distribution, which generates networks with properties similar to those of real road networks~\cite{sohouenou2020using}. All numerical experiments for the St. Thomas data were done on a laptop with a 10 core CPU and 32 GB of RAM, and we used Gurobi 9.5.2 software~\cite{gurobi} to perform our network optimization experiments. For the remainder of the experiments, we utilized machine with a 16 core CPU and 64 GB RAM, running Gurobi 10.0.0.

\begin{table}[h!]
    \centering
    \scalebox{0.68}{\begin{tabular}{c|c|c|c|c|c|c|c}
    \emph{data} & \# nod. & \# ed. & dens. & avg. deg. & deg. het. & max deg. & avg. betw. \\
    \hline
% \textit{USVI} & 224 & 436 & 0.009 & 2.045 & 0.42 & 4 & 0.336\\
% \textit{Anaheim} & 416 & 914 & 0.007 & 3.048 & 0.977 & 7 & 0.118\\
% \textit{Power law}& 49 & 94 & 0.04 & 1.918 & 1.338 & 8 & 0.112\\
% & 100 & 194 & 0.02 & 1.94 & 1.788 & 16 & 0.084\\
% & 150 & 298 & 0.013 & 1.987 & 1.781 & 16 & 0.067\\
% & 200 & 392 & 0.01 & 1.96 & 1.962 & 20 & 0.041\\
% & 250 & 486 & 0.008 & 1.944 & 1.993 & 18 & 0.035\\
% & 300 & 584 & 0.007 & 1.947 & 2.384 & 33 & 0.024\\
% \textit{Power law} & 350 & 680 & 0.006 & 1.943 & 2.435 & 37 & 0.022\\
% \textit{Power law} & 399 & 786 & 0.005 & 1.97 & 2.166 & 24 & 0.03\\
% \textit{Exponential} & 199 & 1034 & 0.026 & 5.196 & 8.233 & 80 & 0.023\\
% \textit{GREREC} & 16 & 52 & 0.217 & 3.25 & 1.031 & 5 & 0.389\\
% & 25 & 82 & 0.137 & 3.28 & 1.281 & 7 & 0.224\\
% & 36 & 142 & 0.113 & 3.944 & 1.999 & 8 & 0.255\\
% & 48 & 130 & 0.058 & 2.708 & 0.978 & 5 & 0.298\\
% & 63 & 190 & 0.049 & 3.016 & 1.485 & 8 & 0.169\\
% & 79 & 300 & 0.049 & 3.797 & 1.679 & 8 & 0.138\\
% & 99 & 330 & 0.034 & 3.333 & 1.393 & 7 & 0.137\\
% & 120 & 424 & 0.03 & 3.533 & 1.581 & 8 & 0.168\\
% & 144 & 508 & 0.025 & 3.528 & 1.645 & 8 & 0.172\\
% & 167 & 594 & 0.021 & 3.557 & 1.558 & 8 & 0.148\\
% & 194 & 652 & 0.017 & 3.361 & 1.524 & 8 & 0.163\\
% \textit{GREREC} & 224 & 748 & 0.015 & 3.339 & 1.536 & 8 & 0.139\\
% & 254 & 886 & 0.014 & 3.488 & 1.594 & 8 & 0.132\\
% & 286 & 1070 & 0.013 & 3.741 & 1.725 & 8 & 0.13\\
% \textit{GREREC} & 323 & 1190 & 0.011 & 3.684 & 1.647 & 8 & 0.128\\
\textit{St. Thomas, USVI} & 224 & 436 & 0.009 & 2.045 & 0.42 & 4 & 0.336\\
\textit{Anaheim} & 416 & 914 & 0.007 & 3.048 & 0.977 & 7 & 0.118\\
\textit{Power law} & 350 & 688 & 0.006 & 1.966 & 2.17 & 19 & 0.046\\
\textit{Exponential} & 199 & 1034 & 0.026 & 5.196 & 8.233 & 80 & 0.023\\
\textit{GREREC}  & 223 & 802 & 0.016 & 3.596 & 1.461 & 8 & 0.165\\
    \hline
    
    \end{tabular}}
    \caption{Summary statistics and network measures for St, Thomas, USVI (fictitious, yet realistic), Anaheim (realistic), as well as for maximal parameters of the three synthetic models (power law, exponential, and GREREC) we used in experiments. In particular, we present the number of nodes, number of edges, density, average degree, degree heterogeneity (standard deviation of the degree distribution), and average betweenness centrality~\cite{freeman2002centrality} (average $\ell_1$-normalized betweenness centrality).}
    \label{tab:statistics}
\end{table}

\subsection{St. Thomas, USVI}
The St. Thomas, USVI data present a fictitious, yet realistic scenario for interdependent transportation and fuel networks on a metropolitan scale (i.e., 32-square miles). The data include information about location of fuel supply (gas depot) and fuel demand (gas station) nodes based on the fuel network model presented in~\cite{alderson2015operational}, the roads linking them, the speed limit and traffic capacity of each road, the fuel capacities at each node, as well as penalties for unmet demand. This fictitious fuel pipeline is linked to real locations for fuel supply (i.e., gas stations) along simplified roadway geometry that matches the real-world road network on St. Thomas.
Road data are based on~\cite{routley2020operational} and are informed by data provided by the St. Thomas, USVI government and associated agencies. In total, there are 2 fuel depot nodes, 12 gas station nodes, and 210 transport junction (non-fuel) nodes, for a total of 224 nodes -- and 436 edges among these nodes. We present a geospatial view of the network in Figure~\ref{fig:usvi}. Summary network statistics are presented in Table~\ref{tab:statistics}. Average degree of around 2 and network density of 0.009 indicates a sparsely connected road network where a node is typically connected to two other nodes along particular paths. This network structure is common for fuel and roadway systems that have limited redundancy such as those found on islands in the Caribbean and Pacific. Average betweenness centrality of 0.336 suggests presence of nodes with many shortest paths passing through them. 
%While the average degree, density, and maximum degree are all very modest, average betweenness centrality is quite high, indicating that .

\subsection{Anaheim}
The Anaheim dataset is a widely-used, realistic transportation use case taken from Transportation Networks for Research~\cite{stabler2018transportation}. The network consists of 416 nodes and 914 edges, representing an area covering roughly 50 square miles. In addition to the network topology, the dataset contains information about speed limits, road length, and geospatial coordinates. We visualize the network using a geospatial view in Figure~\ref{fig:anaheim}. As the dataset lacks information regarding locations of fuel depots and gas stations, we randomly assigned nodes to correspond to depots, stations, and junctions, in proportion roughly similar to those found in the St. Thomas, USVI data. Network statistics and measures are presented in Table~\ref{tab:statistics}. A relatively small density of 0.007 and average degree of around 3 indicate a relatively sparse network, though the maximum degree is quite high at 7, indicating a junction with 7 edges. The average betweenness centrality is quite low, indicating a more homogeneous distribution of important conduit nodes than in the St. Thomas, USVI data.

\subsection{Power law}
\begin{figure*}[tb]
    \centering
    \includegraphics[width = 0.35\linewidth]{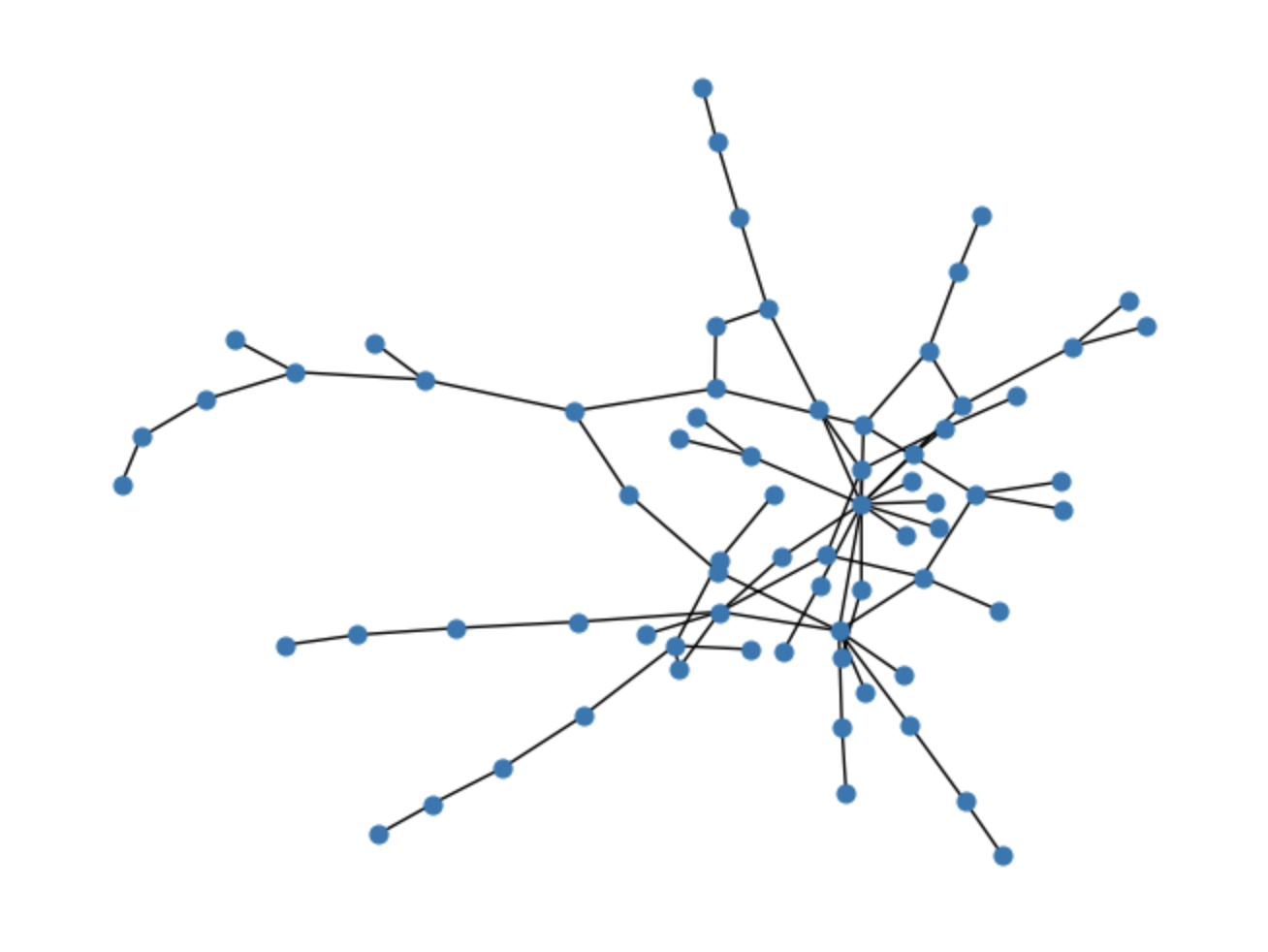}
    \includegraphics[width = 0.35\linewidth]{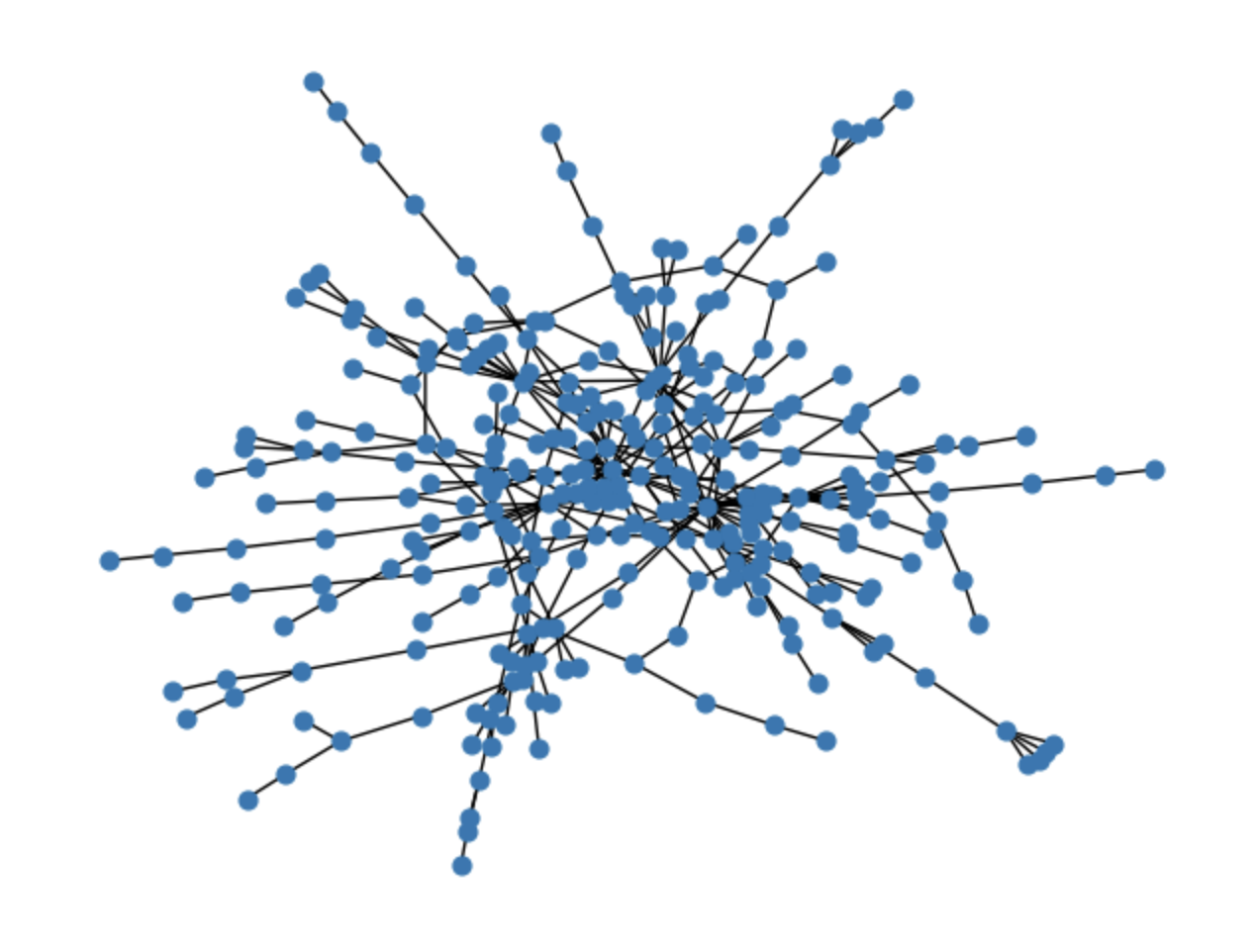}
    \caption{(Left). An instance of a power law network for $n=100,$ $\gamma=3.0$. (Right). An instance of a power law network for $n=400,$ $\gamma=3.0$.}
    \label{fig:power}
\end{figure*}
\begin{figure*}[tb]
    \centering
    \includegraphics[width = 0.35\linewidth]{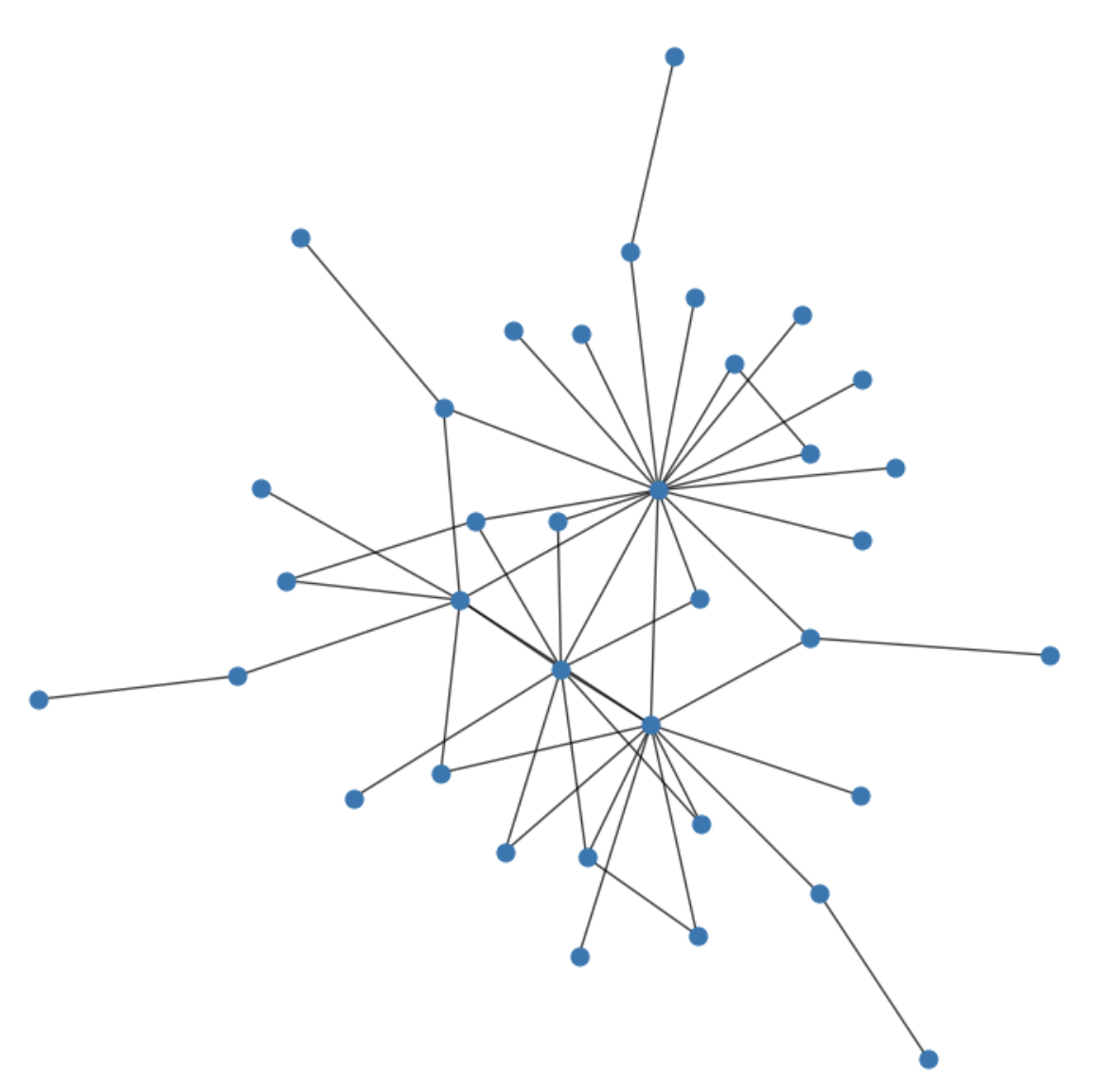}
    \includegraphics[width = 0.35\linewidth]{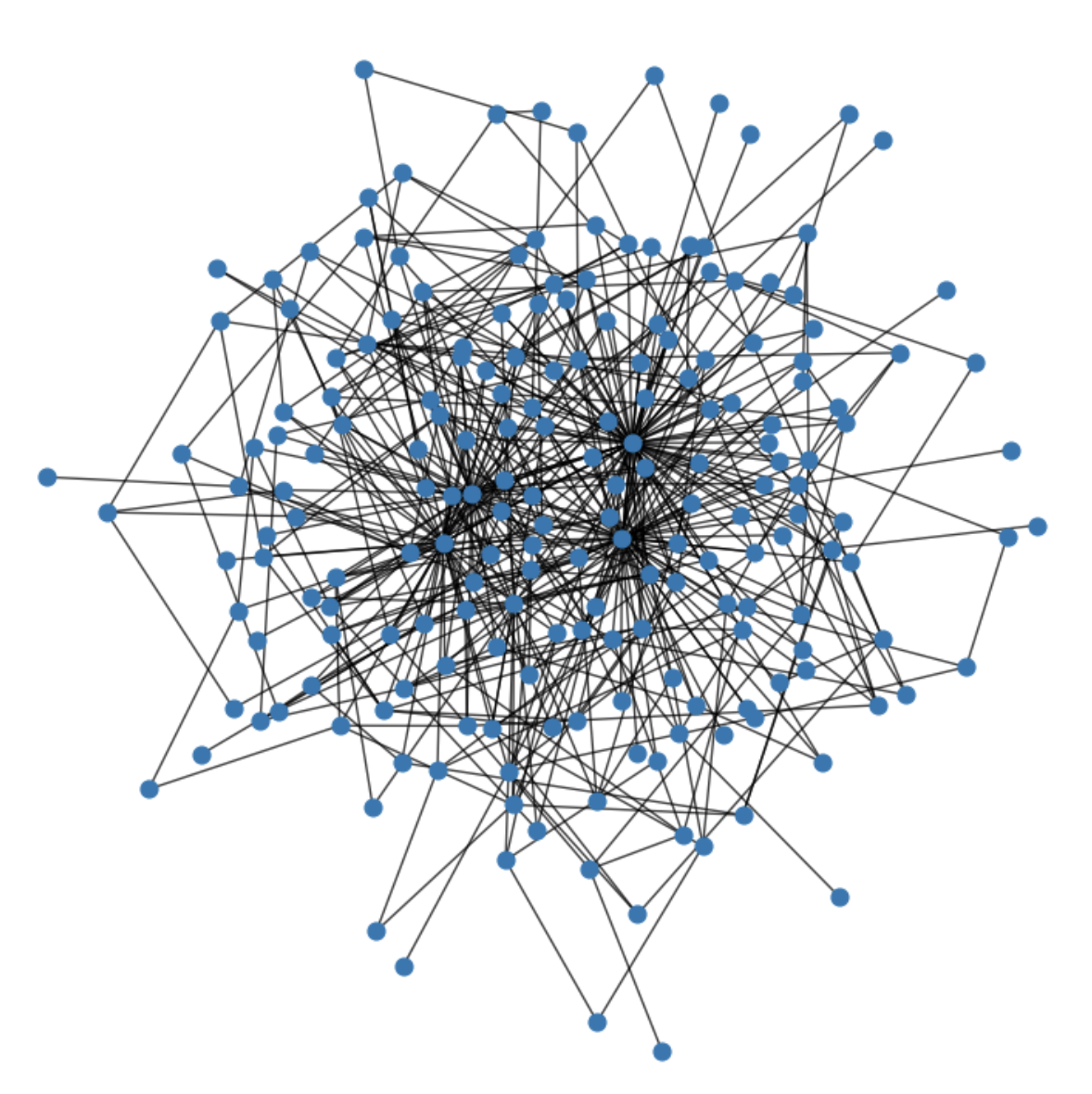}
    \caption{(Left). An instance of an exponential network for $n=50$. (Right). An instance of an exponential network for $n=200$.}
    \label{fig:exponential}
\end{figure*}
\begin{figure*}
    \centering
    \includegraphics[width = 0.4\linewidth]{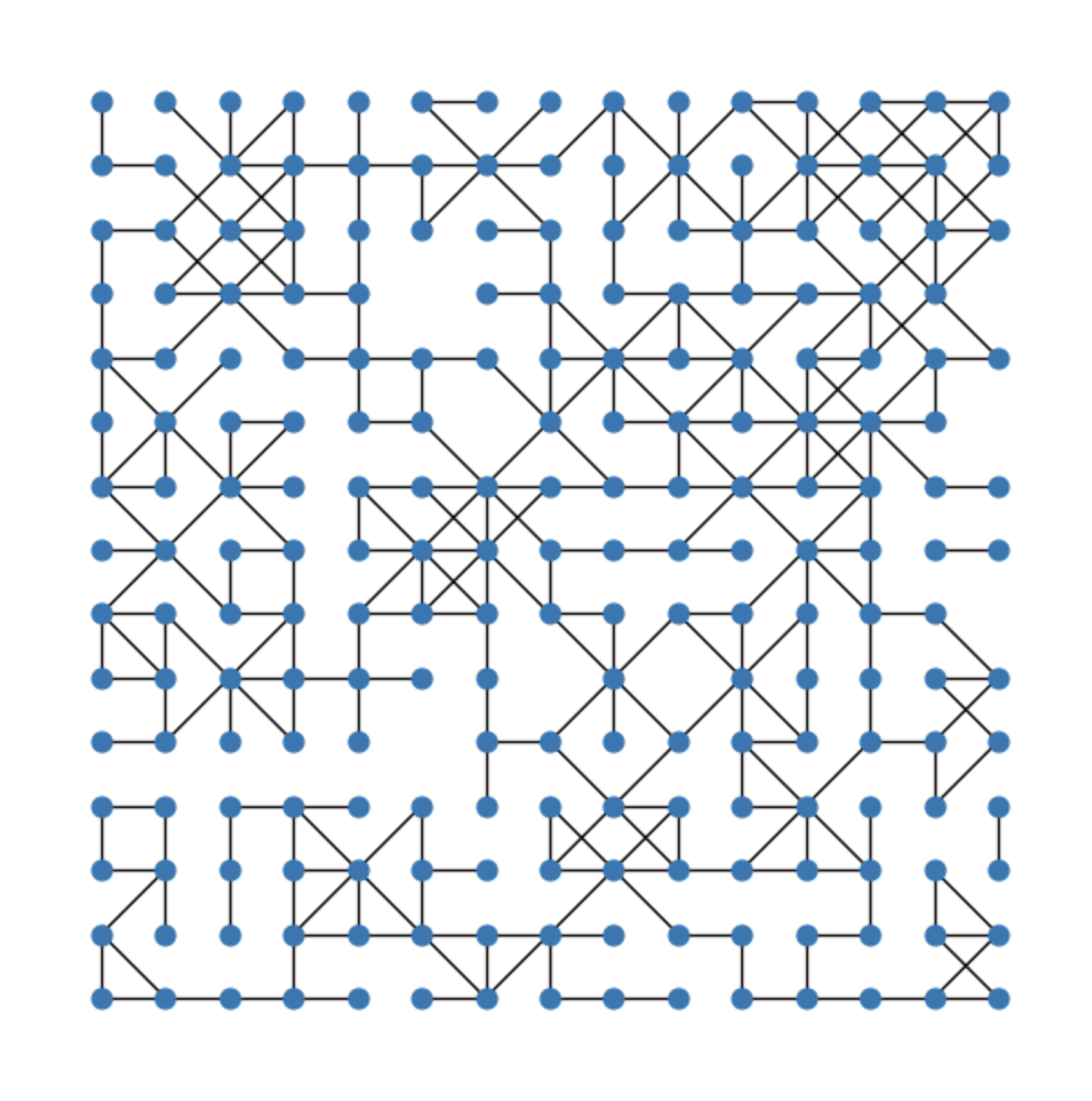}
    \includegraphics[width = 0.4\linewidth]{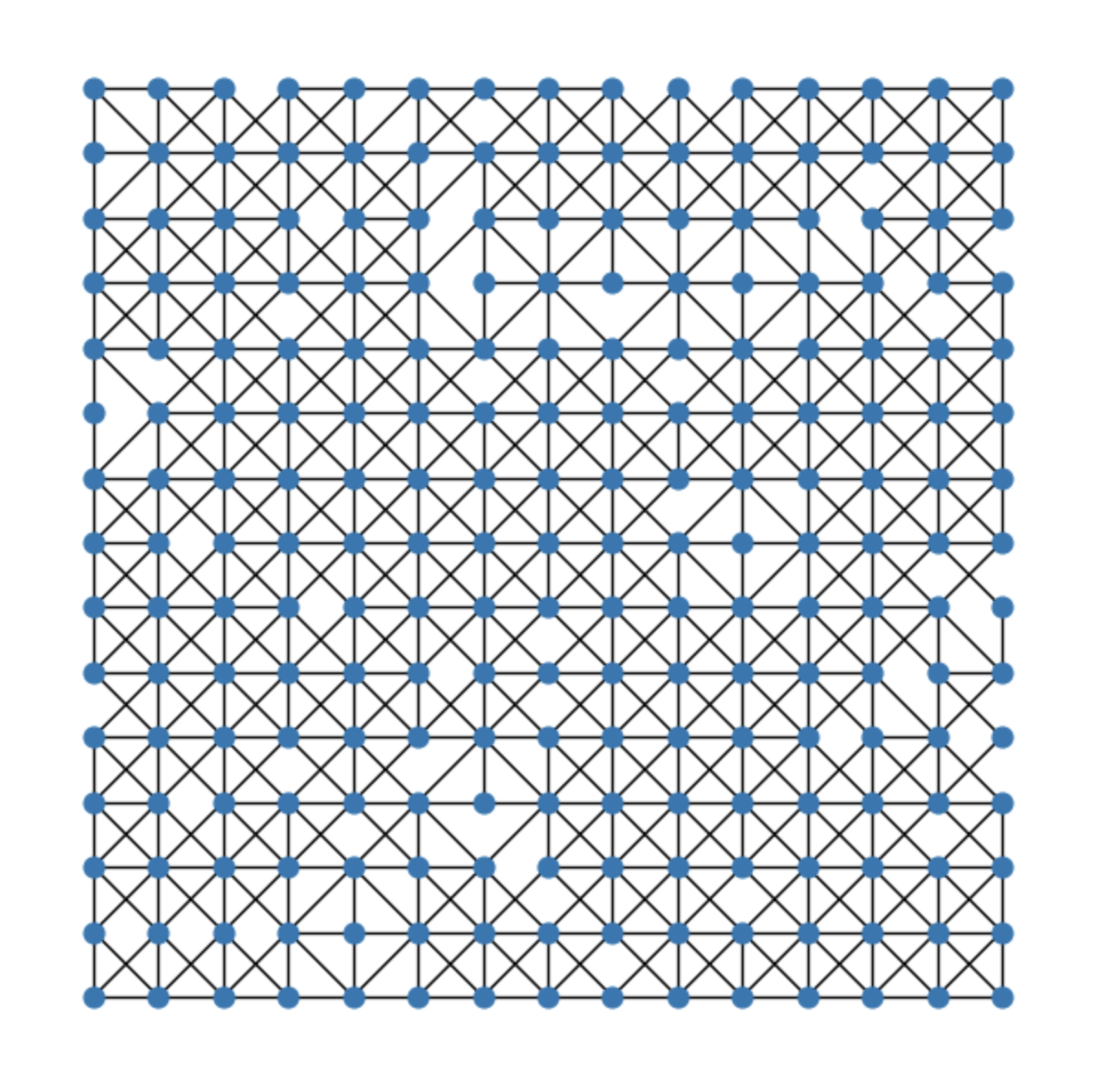}
    \caption{(Left). An instance of a GREREC random road network for $m=n=15,$ $p=0.5,q=0.25$. (Right). An instance of a GREREC random road network for $m=n=15,$ $p=0.9,q=0.9$.}
    \label{fig:grerec}
\end{figure*}
Here, the road network is generated from a power law degree distribution with an exponent of $\gamma=3.0$. Summary network statistics for the largest network we sampled are presented in Table~\ref{tab:statistics}. This network has 350 nodes and 688 edges, where the density and expected degree are both quite low, as characteristic of power law networks with our choice of exponent -- though the maximum degree is quite high, at 19 -- as is the degree heterogeneity, at around 2.2. A very low average betweenness centrality suggests that most nodes are not important conduits.  Two of the networks we sampled are shown in Figure~\ref{fig:power}. Visually, the networks tend to have dense cores and sparse peripheries, reminiscent of cities. We randomly choose approximately 20\% of nodes to be fuel nodes
% in each mode, and overlap 5\% of nodes across modes
\footnote{We do the same for all the random network models we examine in this subsection, unless we specify otherwise.}.

\subsection{Exponential}
In~\cite{zhang2017backbone}, it was observed that exponential degree distributions with degree probability distributions of the form $$p(k) = a_0+\frac{A}{w\sqrt{\pi/2}}\exp{\left[-2\left(\frac{k-k_c}{w^2}\right)^2\right]}$$ fit real road network degree distributions well. We present two networks sampled from this distribution in Figure~\ref{fig:exponential}.

% We run our model on random graphs with degree sequences drawn from this distribution, varying the number of nodes. 
Summary network statistics are reported in Table~\ref{tab:statistics} for the largest network network we sampled from this distribution, which has 199 nodes and 1034 edges. Note that while the betweenness is the lowest among the networks tested, average degree is the highest, and the the maximum degree is a whopping 80, representing a relatively dense network.

\subsection{GREREC}
The GREREC model allows the generation of networks that have topologies and failure characteristics similar to real-world road networks~\cite{sohouenou2020using}. The model allows for four parameters to control the size and density of the sampled networks. In particular, the parameters $m$ and $n$ control the grid with and height, respectively, while the parameters $p$ and $q$ control the probability of keeping a given edge in the grid, and the probability of adding diagonal edges at any given node, respectively. Figure~\ref{fig:grerec} shows two instances drawn from the GREREC model. We record summary statistics for the largest network we sampled from this distribution in Table~\ref{tab:statistics}, which has 223 nodes and 802 edges, and corresponds to parameters $n=m=15,$ and $p=0.7,$ $q=0.2.$ As expected for these parameter ranges, the average degree is around 3.6, and the maximum degree is 8, which indicates a relatively dense network. The network also has the highest betweenness centrality among all the maximal-size synthetic networks we presented here.

\begin{figure}
    \centering
    \includegraphics[width=0.8\linewidth]{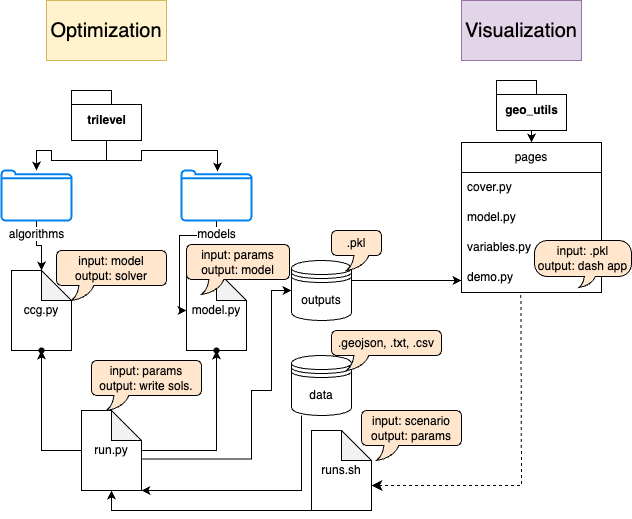}
    \caption{A high-level overview of our codebase.}
    \label{fig:code}
\end{figure}
\subsection{Computational workflow}
%Here, we present an outline of the implementation of our model. A codebase visualization is presented in 
Figure~\ref{fig:code} presents an overview of our codebase for the tri-level optimization model. In particular, we implement the model in Python, using Pyomo to interface with the Gurobi solver. The code is divided across optimization and visualization modules. The optimization module is split up into utility files that facilitate I/O with the input data (runs.sh, run.py), as well as files that form the model in this paper (model.py) and files that implement the CCG algorithm (ccg.py). The visualization module facilitates an interactive Plotly dash app visualization of the results from the optimization module, and we use it to create the plots for the St. Thomas, USVI use case. In total, the codebase has approximately 15,000 lines of code.

\section{Results and Discussion}
In this section, we present results on the generalizability and scalability of our framework. In particular, we demonstrate generalizability by showing the results of running our model on each of the five networks in Table~\ref{tab:statistics} -- namely, which nodes the defender chooses to defend, along with the reserve nodes the defender chooses to open -- in addition to the the attacker's choice. Furthermore, we investigate the scalability of our approach with network size. In particular, to demonstrate scalability, we present runtime results for networks of various sizes sampled from the three synthetic distribution discussed in the previous section.
\subsection{Generalizability}
In this subsection, we demonstrate the ease with which our framework generalizes across networks. In particular, we present the outcome of running our model on the five networks seen in Table~\ref{tab:statistics}. In each case, we were able to successfully solve the DAD model formulation. The results could be used to gain practical, mitigation-relevant insights for each of the networks.
\paragraph{St. Thomas, USVI}

\begin{figure*}[t!]
    \centering
    \includegraphics[width=0.65\linewidth]{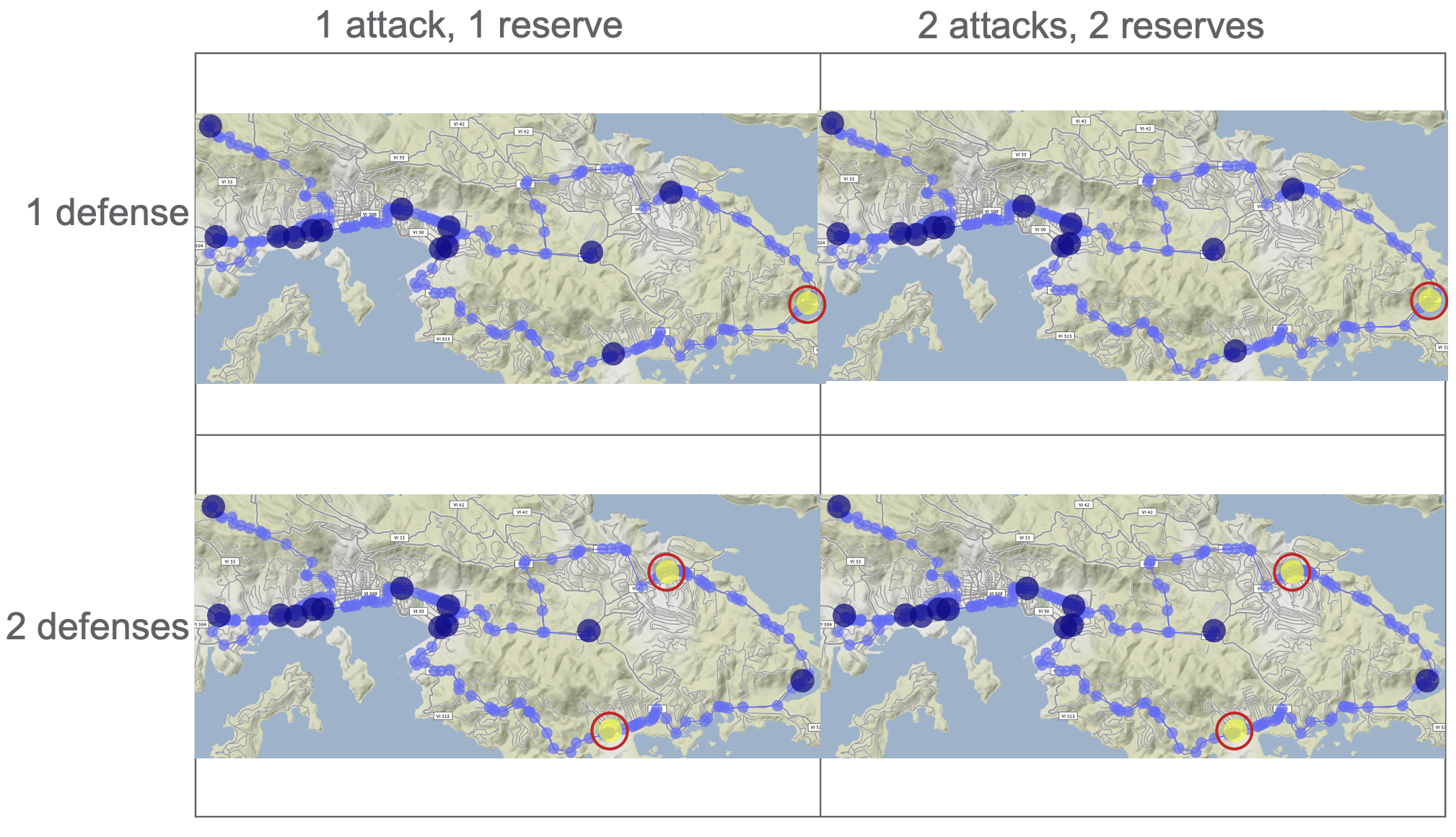}
    \caption{Defender’s protective actions that optimally mitigate fuel shortages. Selected nodes represent 1 or 2 of the 12 gas stations chosen by the defender.}
    \label{fig:defense}
\end{figure*}

\begin{figure*}[h!]
    \centering
    \includegraphics[width=0.65\linewidth]{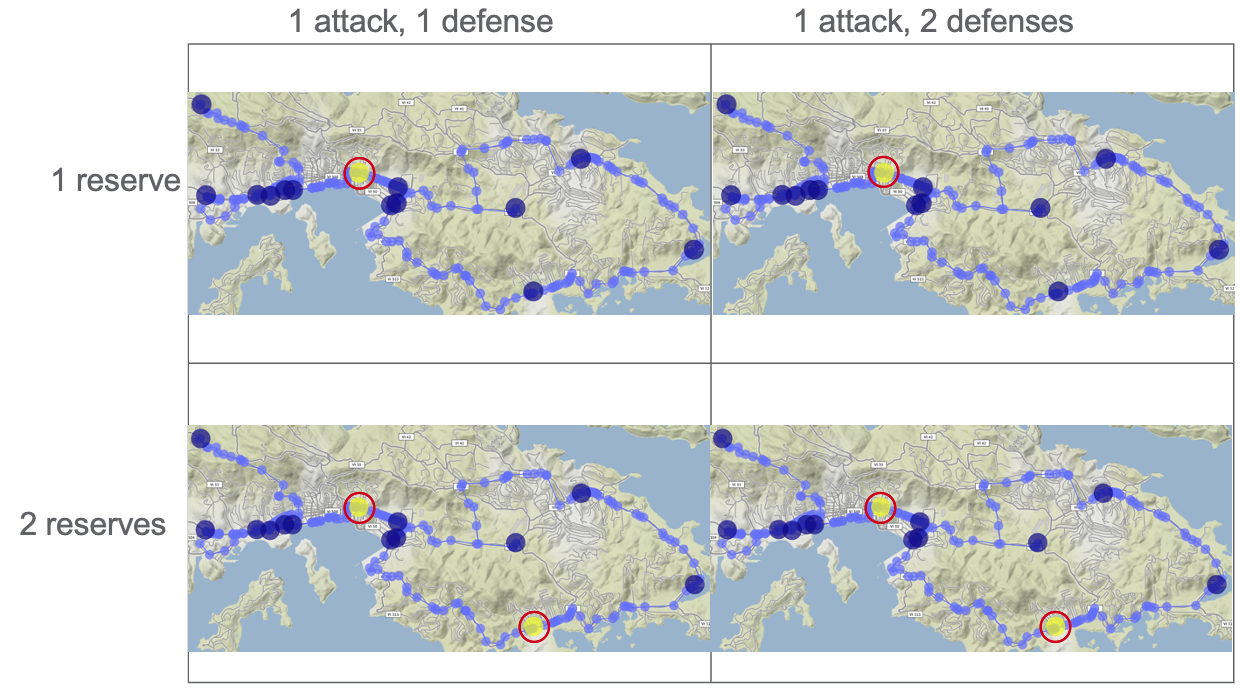}
    \caption{
Defender’s reserve opening choices that optimally mitigate fuel shortages. Selected nodes represent 1 or 2 of the 12 gas stations selected by the defender.
}
    \label{fig:reserve}
\end{figure*}
\begin{table}[]
    \centering
    % \scriptsize
    \begin{tabular}{c|c|c}
         &  1 defense, 1 reserve & 2 defenses, 2 reserves\\
         \hline
        1 attack & $v_1$ & $v_1$ \\
        \hline
        2 attacks & $v_1,v_2$ & $v_1,$ $v_2$
    \end{tabular}
    \caption{Adversary’s choices leading to maximal network supply disruptions. Selected anonymized nodes represent 1 or 2 of the 12 gas stations selected by the adversary.}
    \label{tab:attack}
\end{table}
Figures~\ref{fig:defense},~\ref{fig:reserve}, and Table~\ref{tab:attack} present simulation results from our DAD model using St. Thomas, USVI data for a variety of defense, reserve, and attack budget scenarios described below. Generating solutions from the DAD model took around a minute of computational time for each scenario. We present results from phase 2 of the optimization, where the 12 gas stations act as supply nodes.

In Figure~\ref{fig:defense}, we assign the attack and reserve budgets (1 or 2 nodes each) and vary the number of defenses (1 or 2 nodes), and seek to identify the nodes that a defender might harden. Here, a budget higher than 1 indicates a simultaneous compound attack. The highlighted nodes represent the defender's actions that optimally mitigate fuel shortages. Note that the level of attack and reserve budget does not impact these optimal choices. Next, in Figure~\ref{fig:reserve}, we assign the number of attacks (1 node) and defenses (1 or 2 nodes) and vary the reserve budget (1 or 2 nodes), and seek to identify the reserve nodes that a defender might open. The highlighted nodes correspond to the defender's actions (opening up of reserve nodes) that optimally mitigate fuel shortages. Again, the level of attack and defense budget does not impact these optimal choices. Lastly, in Table~\ref{tab:attack}, we assign the reserve and defense budgets (1 or 2 nodes each) while varying the attack budget (1 or 2 nodes), and seek to identify the nodes that an adversary might target. The identified nodes\footnote{Due to issues of sensitivity, we have anonymized the nodes labels so that the locations of $v_1$ and $v_2$ on the map are hidden.} in this case represent optimal actions, but from the adversary's perspective -- these choices lead to the maximal fuel supply disruptions. As in the last two cases, the level of defense and reserve does not impact these optimal choices. Note that the defender's protective actions, defender's reserve opening actions, and the attacker's choices do not overlap for any given defense, reserve, and attack budget.
\begin{figure*}
    \centering
    \includegraphics[width = 0.7 \linewidth]{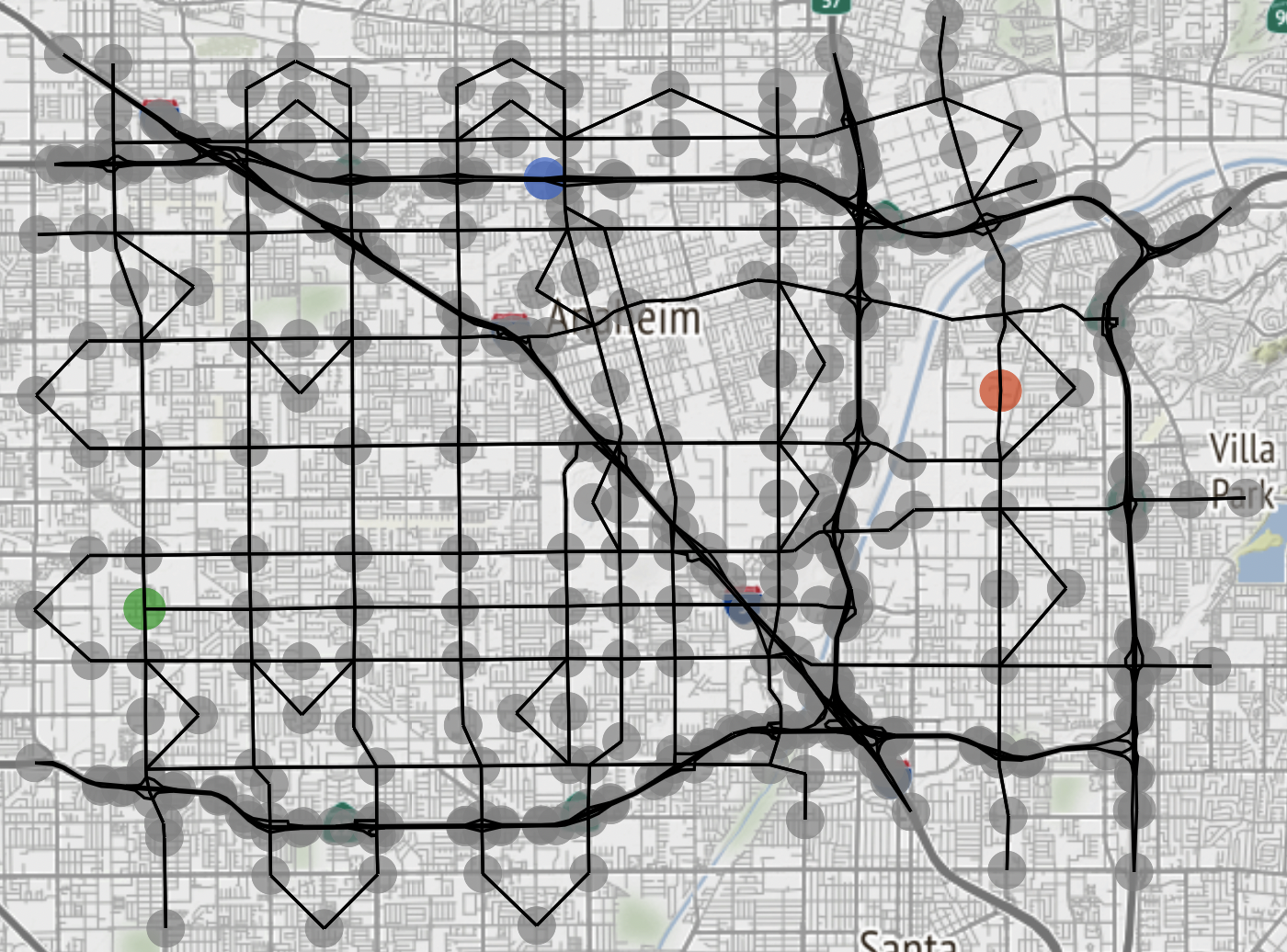}
    \caption{Defense (blue), reserve (green), and attack (red) node choices for the Anaheim transportation network. }
    \label{fig:anaheim_results}
\end{figure*}
\begin{figure*}
    \centering
    \includegraphics[width = 0.7 \linewidth]{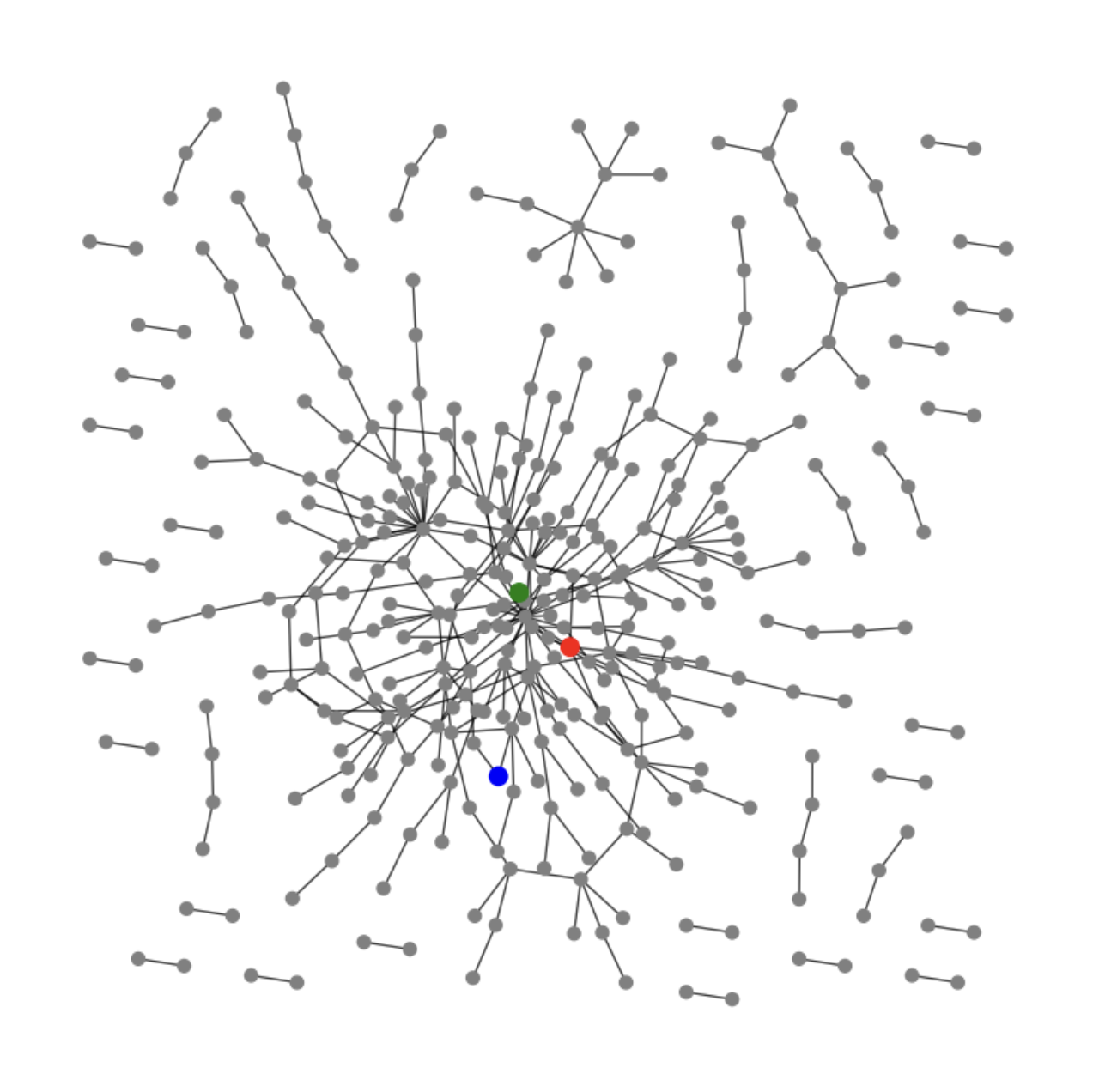}
    \caption{Defense (blue), reserve (green), and attack (red) node choices for a network drawn from the power law degree distribution with $n=350$ and $\gamma = 3.0$.}
    \label{fig:power_results}
\end{figure*}
\begin{figure*}
    \centering
    \includegraphics[width = 0.65 \linewidth]{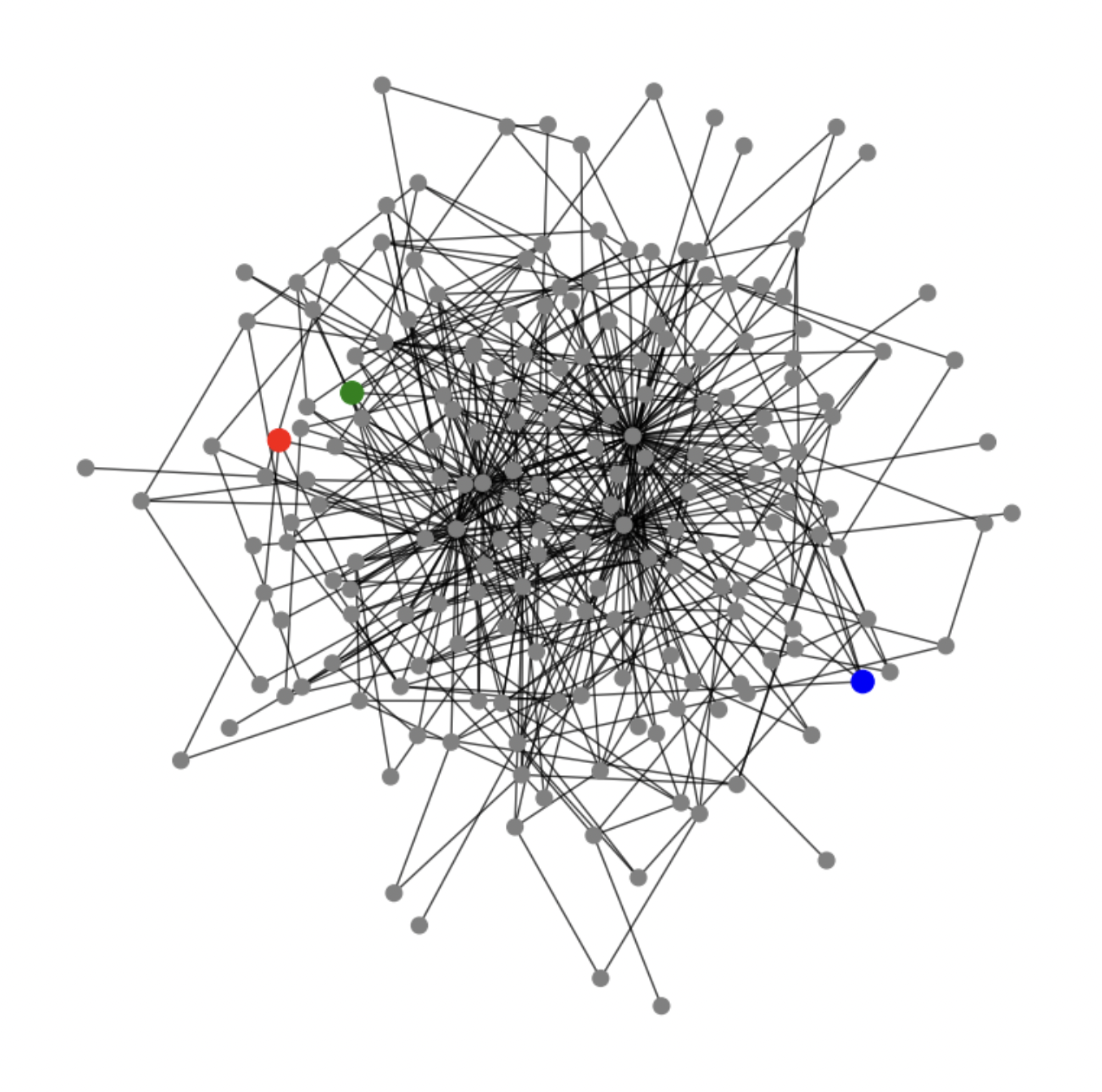}
    \caption{Defense (blue), reserve (green), and attack (red) node choices for a network drawn from the exponential distribution with $n=200$.}
    \label{fig:exponential_results}
\end{figure*}
\begin{figure*}
    \centering
    \includegraphics[width = 0.57 \linewidth]{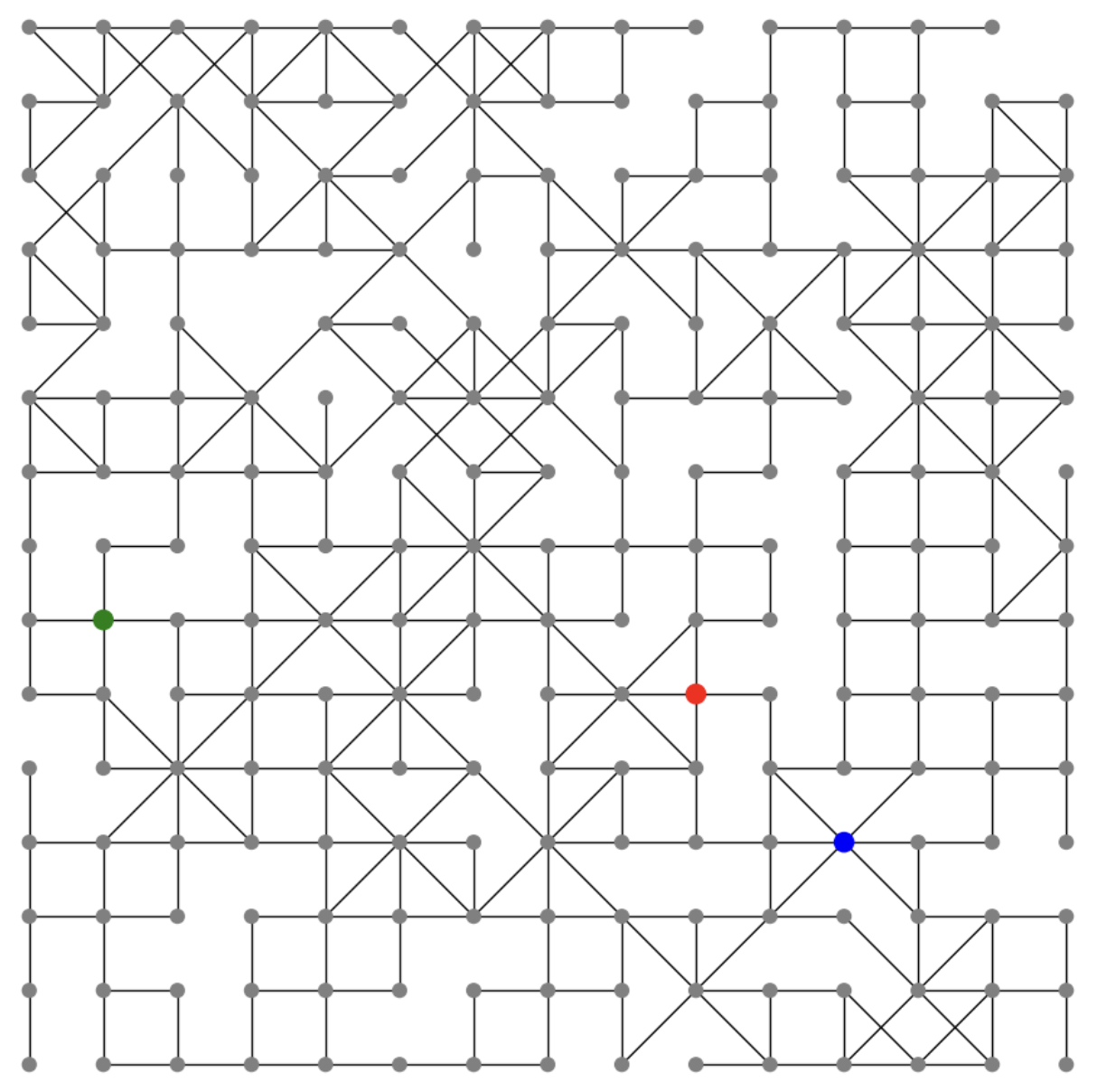}
    \caption{Defense (blue), reserve (green), and attack (red) node choices for a network drawn from the GREREC distribution with $p=0.7, q=0.2$ and $m=n=15$.}
    \label{fig:grerec_results}
\end{figure*}

%\begin{figure*}[h]
%    \centering
%    \includegraphics[width = 0.6 %\linewidth]{images/optimization-time}
%    \caption{The runtimes of our model on synthetic networks drawn from power law, exponential, and GREREC distributions all scale roughly as $O(N^{3.5}),$ where $N$ is the number of nodes.}
%    \label{fig:synth_times}
%\end{figure*}

\begin{figure}[t]
    \centering
    \includegraphics[width=0.99\linewidth]{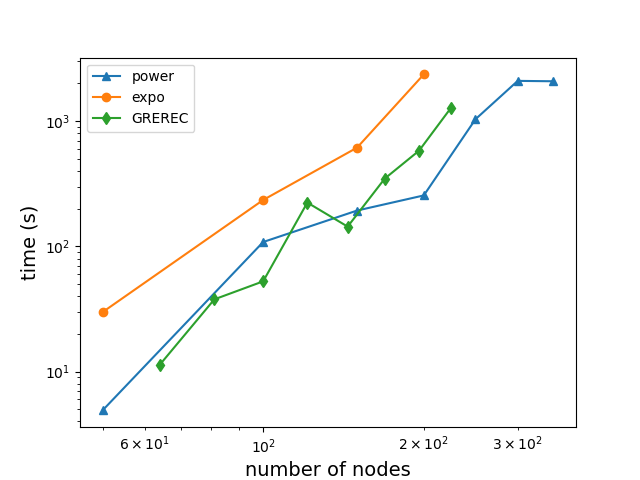}
    \caption{The runtimes of our model on synthetic networks drawn from power law, exponential, and GREREC distributions all scale roughly as $O(N^{3.5}),$ where $N$ is the number of nodes.}
    \label{fig:synth_times}
\end{figure}

%\begin{figure*}[!b]
%    \centering
%    \includegraphics[width = 0.60\linewidth]{images/grerec_time}
%    \caption{Our model runs on even dense (high $p$ and $q$) instances of the GREREC random road networks in under 3 minutes.}
%    \label{fig:grerec_times}
%\end{figure*}

\begin{figure}
    \centering
    \includegraphics[width=0.99\linewidth]{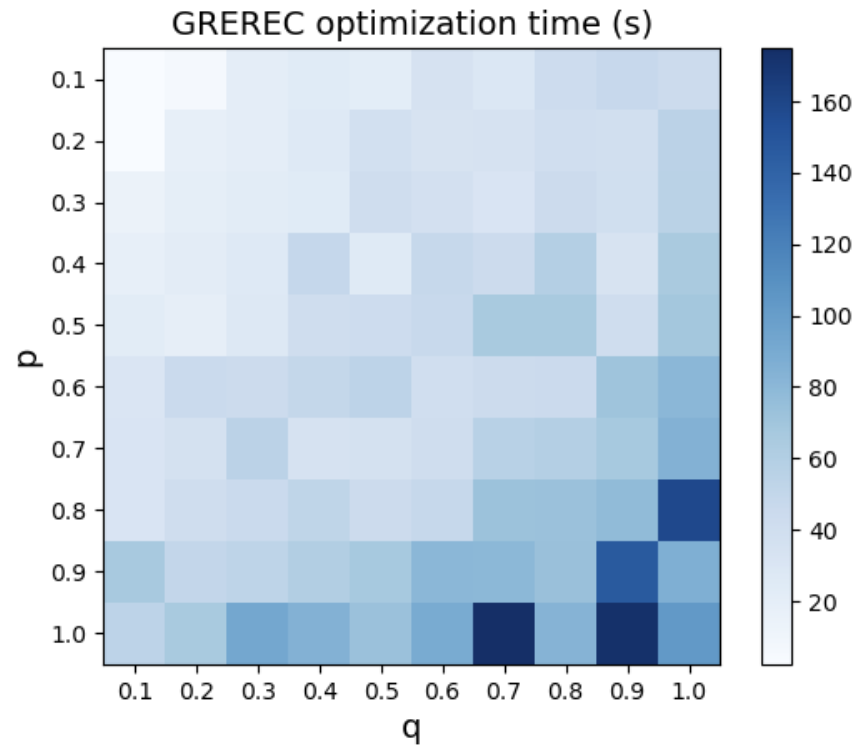}
    \caption{Our model runs on even dense (high $p$ and $q$) instances of the GREREC random road networks in under 3 minutes.}
    \label{fig:grerec_times}
\end{figure}

\paragraph{Anaheim}
% \begin{table*}[t]
%     \centering
%     \scalebox{0.8}{\begin{tabular}{c|c|c|c}
%     \emph{dataset} & node defended & reserve opened & node attacked \\
%     \hline
% \textit{USVI} & 224 & 436 & 0.009\\
% \textit{Anaheim} & 416 & 914 & 0.007\\
% \textit{Power law} & 399 & 786 & 0.005\\
% \textit{Exponential} & 199 & 1034 & 0.026\\
% \textit{GREREC} & 323 & 1190 & 0.011\\
%     \hline
    
%     \end{tabular}}
%     \caption{Defender and attacker choices for the realistic Anaheim network, and three synthetic networks (power law, exponential, and GREREC) given a budget of 1 node to defend, 1 reserve to open, and 1 node to attack.}
%     \label{tab:results}
% \end{table*}
With a defense, reserve, and attack budget of 1, we were able to run our model on this dataset in around 20 hours. This result illustrates the ability of our model to scale up to the level of cities, and to generalize between settings. We show the result of running our model on this dataset in Figure~\ref{fig:anaheim_results}. The node highlighted in blue represents the defender actions (node hardening) that optimally mitigate fuel shortages, while the node highlighted in green represents the defender's reserve opening choice that leads to optimal fuel shortage mitigation -- and lastly, the node in red represents the adversary choice leading to maximal fuel supply disruptions.

\paragraph{Power law}
With a defense, reserve, and attack budget of 1, we ran our model on an instance drawn from the power law degree distribution with $N=350$ and $\gamma = 3.0$ in under an hour. We show the node that was chosen to harden by the defender (blue), the reserve node that was opened by the defender (green), and the node that was attacked by the adversary (red) in Figure~\ref{fig:power_results}.

\paragraph{Exponential}
With a defense, reserve, and attack budget of 1, we were able to run our model on a network drawn from the exponential distribution with $N = 200$ in around an hour. We show the node that was chosen to defend (blue), the reserve node that was opened (green), and the node that was attacked (red) in Figure~\ref{fig:exponential_results}.

\paragraph{GREREC}
With a defense, reserve, and attack budget of 1, we ran our model on this dataset in under an hour. We show the node that was chosen to harden by the defender (blue), the reserve node that was opened by the defender (green), and the node that was attacked by the adversary (red) in Figure~\ref{fig:grerec_results}.

\subsection{Scalability}
Here, we demonstrate scalability of our model with the size of the network. In particular, we run our model on networks of increasing size drawn from power law, exponential, and GREREC distributions.

\paragraph{Power law}
The optimization times as a function of the number of nodes $N$ are presented in Figure~\ref{fig:synth_times}. Fitting to a power law, we see that the runtime scales roughly as $O(N^{3.05})$.

\paragraph{Exponential}
 We report runtimes of our model on instances drawn from the exponential distribution in Figure~\ref{fig:synth_times}. Unlike the previous networks, the results presented here correspond to running our model with two modes. Again fitting to a power law, we see that the runtime scales roughly as $O(N^{3.04})$, where $N$ is the number of nodes. This is quite impressive, given that we actually ran the model on two overlapping networks (modes) for each run, with around 5\% of nodes overlapped.

\paragraph{GREREC}
Figure~\ref{fig:grerec_times} presents the runtimes of our model on networks generated with $n,m=10$ over a range of $p$ and $q$ probabilities.
Next, we investigate the way in which the runtime scales with the grid size. Figure~\ref{fig:synth_times} (middle) presents the results for fixed $p=0.7$ and $q=0.2$ and varying grid width and height, where we set $m=n$. We report the total number of nodes ($N\equiv mn=m^2$) as the independent variable. Fitting to a power law, we see that the runtime scales roughly as $O(N^{3.42})$.

\subsection{Discussion}
The methodology and results in this paper represent a computational framework for generating practical mitigation insights associated with interdependent networked critical infrastructures in the presence of compound hazards under budget constraints. While the focus was on an interdependent fuel and transportation system, the approach can generalize to other interdependent systems. Attack choices represent failure conditions which may lead to maximal transportation network fuel supply disruptions. On the other hand, defense choices represent protective strategies that may lead to optimal mitigation of fuel shortages. These results may provide practical mitigation-relevant insights for wargaming exercises and what-if scenario analyses.

% %%% FUTURE WORK %%%
\section{Conclusions and Future Work}

In this paper, we presented a tri-level DAD model for analyzing defense, operator, and attack scenarios for interdependent infrastructure networks in the presence of compound attack events. We implemented the DAD model to analyze interdependent fuel and transportation networks. However, it is crucial to note that the model formulation can accommodate other interdependent critical infrastructure scenarios. Modeling results are in the form of practical mitigation options for the defender. In other words, for any instance of our DAD model, the solution is a plan or mitigation strategy that indicates how to best prepare for consequences on the system under impact. A portfolio of such outcomes can be gathered by performing a sensitivity analysis, by changing hazard event scenarios or defender policies. We demonstrated the generalizability and effectiveness of our model by applying it to the fictitious, yet realistic interdependent fuel and transportation network of St. Thomas, USVI and the realistic Anaheim network, as well as networks generated from three synthetic distributions -- yielding computationally feasible practical results and defender mitigation and response planning insights. In addition, we demonstrated the scalability of our model on networks of various sizes drawn from the three synthetic distributions.

In the future, we plan to investigate the use of active constraint set learning~\cite{misra2018} to further scale up our implementation. This would enable exploration of larger network and budget sizes. We also plan to expand our modeling formulation to include multistage attacks and defenses, thereby incorporating %generalization would add a 
temporal aspects of consequence preparedness assessments.%, making the model even more realistic.

\section{Acknowledgments}
Pacific Northwest National Laboratory (PNNL) is a multiprogram laboratory operated by Battelle Memorial Institute for the U.S. Department of Energy under Contract No. DE-AC05-76RL01830.
%\clearpage

% %% The Appendices part is started with the command \appendix;
% %% appendix sections are then done as normal sections
% \appendix

% \section{Sample Appendix Section}
% \label{sec:sample:appendix}
% Lorem ipsum dolor sit amet, consectetur adipiscing elit, sed do eiusmod tempor section \ref{sec:sample1} incididunt ut labore et dolore magna aliqua. Ut enim ad minim veniam, quis nostrud exercitation ullamco laboris nisi ut aliquip ex ea commodo consequat. Duis aute irure dolor in reprehenderit in voluptate velit esse cillum dolore eu fugiat nulla pariatur. Excepteur sint occaecat cupidatat non proident, sunt in culpa qui officia deserunt mollit anim id est laborum.

%% If you have bibdatabase file and want bibtex to generate the
%% bibitems, please use
%%
 \bibliographystyle{elsarticle-num} 
 \bibliography{references}%{cas-refs}
 
% \renewcommand{\refname}{}
% \bibliographystyle{plain}\vspace{-2\baselineskip} %plain, unsrt, ieeetr
% \bibliography{references}

%% else use the following coding to input the bibitems directly in the
%% TeX file.

% \begin{thebibliography}{00}

% %% \bibitem{label}
% %% Text of bibliographic item

% \bibitem{}

% \end{thebibliography}
\end{document}